\RequirePackage[]{graphicx} % Loaded here to show images in draft mode, too
\graphicspath{{./Figures/}{./Graphs/}{./Drawings/}}

\documentclass[]{arxsub}

\usepackage[utf8]{inputenc}
\usepackage{graphicx}
\usepackage{environ}
\usepackage{listings}
\usepackage{booktabs}
\usepackage{blindtext}
\usepackage{url}
\usepackage{siunitx}

\usepackage[english]{babel}
\usepackage{csquotes}
\usepackage[
  backend=biber,
  natbib = true,
  style=ext-numeric-comp,
  sorting=none,
  minbibnames=3, maxbibnames=6,
  uniquename=false, uniquelist=false,
  giveninits=true, terseinits,
  articlein=false, innamebeforetitle=true,
  isbn=false,
  urldate=long, dateabbrev=false,
  autocite=superscript,
]{biblatex}

\DeclareNameAlias{sortname}{family-given}

\DeclareNameAlias{author}{sortname}
\DeclareNameAlias{editor}{sortname}
\DeclareNameAlias{translator}{sortname}

\DeclareDelimAlias{finalnamedelim}{multinamedelim}

\DeclareFieldFormat
  [article,inbook,incollection,inproceedings,patent,thesis,unpublished]
  {title}{#1}

\renewbibmacro*{journal+issuetitle}{%
  \usebibmacro{journal}%
  \setunit*{\addspace}%
  \iffieldundef{series}
    {}
    {\newunit
     \printfield{series}%
     \setunit{\addspace}}%
  \usebibmacro{issue+date}%
  \setunit{\addsemicolon\addspace}%
  \usebibmacro{volume+number+eid}%
  \setunit{\addcolon\space}%
  \usebibmacro{issue}}

\DeclareFieldFormat[article,periodical]{number}{\mkbibparens{#1}}

\renewbibmacro*{issue+date}{%
  \usebibmacro{date}}

\renewbibmacro*{pubinstorg+location+date}[1]{%
  \printlist{location}%
  \iflistundef{#1}
    {\setunit*{\addsemicolon\space}}
    {\setunit*{\addcolon\space}}%
  \printlist{#1}%
  \setunit*{\addsemicolon\space}%
  \usebibmacro{date}%
  \newunit}

\DeclareFieldFormat{pages}{#1}

\DeclareFieldFormat{doi}{%
  doi\addcolon\space
  \ifhyperref
    {\href{https://doi.org/#1}{\nolinkurl{#1}}}
    {\nolinkurl{#1}}}

\DeclareFieldFormat{language}{}% nothing
\DeclareFieldFormat{origlanguage}{}%

\AtEveryBibitem{%
\clearlist{language}%
\clearfield{number}%
\clearfield{month}%
\clearfield{day}%
\clearfield{urlyear}%
\clearfield{urlmonth}%
\clearfield{issn}%
\clearfield{series}%
\clearfield{pagetotal}%
\clearfield{note}%
\clearfield{language}%
\clearfield{origlanguage}%
\clearfield{langid}%
\clearfield{eprintclass}%
\ifentrytype{article}{
	\iffieldundef{volume}{}{\clearfield{doi}}%
	\clearfield{eprint}%
	\clearfield{url}%
    \clearfield{publisher}%
}{}%
\ifentrytype{book}{%
	\clearfield{eprint}%
	\clearfield{doi}%
	\clearfield{url}%
}{% Remove publisher and editor except for books
	\clearname{editor}%
}
\ifentrytype{inproceedings}{%
	\clearfield{eprint}%
	\clearfield{url}%
	\clearfield{doi}%
	\clearfield{publisher}%
	\clearlist{publisher}%
	\iffieldundef{booktitle}{}{\clearfield{eventtitle}}%
}{}%
\iffieldundef{eprint}{}{\clearfield{url}}
\iffieldundef{doi}{}{\clearfield{url}}
}

\DeclareSIUnit{\pixel}{px}

\newcommand{\figref}[2][]{\hyperref[#2]{Figure~\ref*{#2}#1}}
\newcommand{\tblref}[2][]{\hyperref[#2]{Table~\ref*{#2}#1}}
\renewcommand{\eqref}[1]{\hyperref[#1]{Eq.~\ref*{#1}}}

\newcommand{\bvec}[1]{\mathbf{#1}}

\providecommand{\doi}{}
\renewcommand{\doi}[1]{\href{https://doi.org/#1}{#1}}

\newcommand{\preSubmission}{}

\providecommand{\del}[2][]{}
\providecommand{\dels}[2][]{}
\providecommand{\printfunding}{}

\title{Fast and Robust Diffusion Posterior Sampling for MR Image Reconstruction Using the
Preconditioned Unadjusted Langevin Algorithm}

\corres{Martin Uecker, Graz University of Technology,
Institute of Biomedical Imaging, Stremayrgasse~16/3, 8010~Graz, AUSTRIA, \email{uecker@tugraz.at}}

\author[1,2]{Moritz Blumenthal$^*$}{}
\author[1]{Tina Holliber\thanks{These authors contributed equally to this work.}}{}
\author[3,4]{Jonathan I.\ Tamir}{}
\author[1,5]{Martin Uecker}{}
\authormark{T. Holliber \textit{et al.}}

\address[1]{Institute of Biomedical Imaging, Graz University of Technology, Graz, Austria}
\address[2]{Department of Radiology, Boston Children's Hospital, Harvard Medical School, Boston, USA}
\address[3]{Chandra Family Department of Electrical Engineering, University of Texas at Austin, USA}
\address[4]{Department of Diagnostic Medicine, Dell Medical School, University of Texas at Austin, USA}
\address[5]{BioTechMed-Graz, Graz, Austria}


\articletype{Technical Note}

\finfo{
The research was funded in whole or in part by the Austrian Science Fund (FWF) 10.55776/F100800 and National Science Foundation (NSF) CCF-2239687 (CAREER).
}

\abstract{
\section{Purpose}
The Unadjusted Langevin Algorithm (ULA) in combination with diffusion models can generate high quality MRI reconstructions with uncertainty estimation from highly undersampled k-space data. However, sampling methods such as diffusion posterior sampling (DPS) or likelihood annealing suffer from long reconstruction times and the need for parameter tuning. The purpose of this work is to develop a robust sampling algorithm with fast convergence.
\section{Theory and Methods}
In the reverse diffusion process used
for sampling the posterior, the exact likelihood is multiplied with the diffused
prior at all noise scales. To overcome the issue of slow convergence, preconditioning is used. The method is trained on fastMRI data and tested on retrospectively undersampled brain data of a healthy volunteer.
\section{Results}
For posterior sampling in Cartesian and non-Cartesian accelerated MRI the new approach outperforms annealed sampling and DPS in terms of reconstruction speed and sample quality.
\section{Conclusion}
The proposed exact likelihood with preconditioning enables rapid and reliable posterior sampling across various MRI reconstruction tasks without the need for parameter tuning.
}

\makeatletter
\patchcmd{\@maketitle}{\vskip 1.5em}{\vskip -0.5em}{}{}
\patchcmd{\@maketitle}{\vskip 1em}{\vskip -0.5em}{}{}
\makeatother

\begin{document}

\maketitle

\section{Introduction}
\label{sec:introduction}
In recent years, diffusion models have shown impressive results in image generation by providing an approach for sampling from a high-dimensional probability distribution.
In a Bayesian setting, they can be used as a learned prior to perform MRI reconstruction even from highly undersampled k-space data by sampling the posterior probability distribution \cite{Jalal_NIPS_2021,Chung_Med.ImageAnal._2022,Luo_Magn.Reson.Med._2023}.
This Bayesian approach to image reconstruction has several key advantages over other deep learning methods: (i) by decoupling the measurement model from the prior, no retraining is necessary when the measurement operator changes (e.g. due to different sampling trajectories, coil arrays, motion, field inhomogeneities); (ii) the measurement noise level uniquely
determines the relative weighting of prior and likelihood, i.e. no tuning of
a regularization parameter is required; and (iii) uncertainty maps computed based on the full posterior can show when the reconstruction becomes unreliable due to an insufficient amount of data \cite{Luo_Magn.Reson.Med._2023}.

While sampling the unconditional prior can be done efficiently using a reverse diffusion process based on a learned series of smoothed prior distributions \cite{Song_Adv.NeuralInf.Process.Syst._2019,Ho_NIPS_2020,Karras_Adv.NeuralInf.Process.Syst._2022}, efficient sampling of the posterior distribution based on a realistic acquisition model such as SENSE \cite{Pruessmann_Magn.Reson.Med._1999} is still a challenging problem. Directly learning the posterior \cite{Gungor__2024}, or various modifications of the likelihood term were proposed \cite{Daras_arxiv_2024,Chung__2025} to improve the sampling of the posterior distribution in a reverse diffusion process, including annealed \cite{Jalal_NIPS_2021} or noisy \cite{Chung_Elev.Int.Conf.Learn.Represent._2023,Janati_Phil.Trans.R.Soc.A._2025} likelihoods.
These methods are relatively slow and require a careful choice of step size and reverse-diffusion noise schedule to achieve good results.
Crucially, these parameters must be tuned for different measurement models, thus losing some of the purported flexibility offered by the Bayesian approach.

For example, \citeauthor{Chung_Elev.Int.Conf.Learn.Represent._2023} showed that using the correct measurement noise level for their DPS \cite{Chung_Elev.Int.Conf.Learn.Represent._2023} method yields poor results. To achieve good reconstruction results, they introduced a heuristic for reweighting the likelihood and prior, which requires tuning of an additional hyperparameter for different measurement models or noise levels and ultimately samples from a distribution that does not correspond to the true posterior.

% putting this in comment for you to decide to expand or to exclude
% A key example elucidating these issues is non-Cartesian sampling, where the highly non-uniform sampling of k-space leads to slow convergence. As we show, ...

In this work, we observe that the practical difficulties encountered when sampling the true posterior distribution are caused by the ill-conditioning of the problem. For convergence, small step sizes and many noise scales must be used which then prevents effective sampling.  We tackle this problem by preconditioning: Degrees of freedom that correspond to large singular values of the SENSE model, i.e. which are strongly restricted by the measurements, are refined with small updates, while degrees of freedom corresponding to small singular values are refined with large updates allowing them to traverse the probability mass given by the learned prior distribution more freely. Our approach can then be used with a fixed pre-chosen step size and thus eliminate the need to tune it for different acquisitions.
The weighting of the likelihood and prior is completely determined by the noise level of the measurements, which is known from an adjustment scan or can be estimated from the image background.

In addition, we show that our method yields faster convergence compared to the annealed likelihood \cite{Jalal_NIPS_2021} and DPS \cite{Chung_Elev.Int.Conf.Learn.Represent._2023} approach.
In particular, this affects non-Cartesian sampling, where the highly non-uniform sampling of k-space leads to a high condition number of the SENSE model and, hence, to slow convergence.
This is issue is partially addressed by the Decomposed Diffusion Sampler (DDS) \cite{Chung_ICLR_2024}, which performs sampling efficiently in Krylow subspaces, but introduces a new hyperparameter for the relative weighting of the likelihood and prior.

We tested the proposed method on both Cartesian and non-Cartesian radial brain data, showing consistently better reconstruction quality and lower computation time compared to annealing and DPS without any need for parameter tuning.
  %newlines are needed

\section{Theory}
\label{sec:theory}

\subsection{Bayesian Reconstruction}
MRI reconstruction can be formulated as the inverse problem
\begin{align}\mathbf{y}=A\mathbf{x}+\mathbf{n},&&\mathbf{n}\sim\mathbb{C}\mathcal{N}(\mathbf{0},\mathbb{I})\,\,\end{align}
with k-space data $\mathbf{y}$, the SENSE model $A$, image $\mathbf{x}$ and white complex Gaussian noise $\mathbf{n}$ of unit variance (after prewhitening, normalization, and adapting the coil sensitivities accordingly).
From the Bayesian perspective, the posterior probability density $p(\mathbf{x}|\mathbf{y})$ is given by
\begin{align}
    p(\mathbf{x}|\mathbf{y}) = \frac{p(\mathbf{y}|\mathbf{x})p(\mathbf{x})}{p(\mathbf{y})}\propto\exp\left(-\lVert\mathbf{y}-A\mathbf{x}\rVert_2^2 + \log p(\mathbf{x})\right) \,, \label{eq:posterior}
\end{align}
with the prior $p(\mathbf{x})$ and the likelihood  $p(\mathbf{y} | \mathbf{x})$. From the posterior distribution, point estimators such as the maximum a posteriori (MAP) estimate, which  corresponds to conventional image reconstruction with regularization, or the minimum mean square error (MMSE) estimate can be computed.
To estimate the MMSE, multiple samples can be drawn from the posterior using the unadjusted Langevin algorithm (ULA) and averaged. For complex valued random vectors, the ULA update reads
\begin{align}
    \mathbf{x}^{k+1}=\mathbf{x}^k+\gamma
    \nabla_{\bar{\mathbf{x}}}\log p(\mathbf{x}^k|\mathbf{y})+\sqrt{2\gamma}\mathbf{z}^k&&\mathbf{z}^k\sim\mathbb{C}\mathcal{N}(\mathbf{0},\mathbb{I})\,,
\end{align}
where $\nabla_{\bar{\mathbf{x}}}\log p(\mathbf{x}|\mathbf{y})$ is the score function of the posterior distribution and $\nabla_{\bar{\mathbf{x}}}$ is the complex conjugate gradient operator of Wirtinger calculus \cite{Kreutz-Delgado__2009}.
The corresponding complex likelihood score is given by
\begin{align}
    \nabla_{\bar{\mathbf{x}}}\log p(\mathbf{y}|\mathbf{x}) = A^H(\mathbf{y}-A\bvec{x})~.
\end{align}
For convergence and to reduce the discretization bias, the step size $\gamma$ must be sufficiently small compared to the inverse Lipschitz $L^{-1}$ constant of the posterior score \cite{Holliber_Proc.Annu.Meet.ISMRM_2025,Dalalyan_J.R.Stat.Soc.Ser.BStat.Methodol._2017}.
The score $\nabla_{\bar{\mathbf{x}}}\log p(\mathbf{x}_0)$ of a smoothed version of the prior can be obtained from a training data set of images using
denoising score matching (DSM) \cite{Song_Adv.NeuralInf.Process.Syst._2019}.

\subsection{Posterior Sampling}

The goal of posterior sampling is to draw samples from the posterior distribution in Eq.~\ref{eq:posterior}, where the prior distribution $p(\mathbf{ x}_0)$ is learned. Because direct sampling of a high-dimensional multi-modal distribution is not practically possible, a reverse diffusion process is used. For sampling of the prior \cite{Song_Adv.NeuralInf.Process.Syst._2019,Ho_NIPS_2020}, this can be done by successive sampling of a series of priors $p(\mathbf{ x}_t)$ smoothed by convolutions with complex Gaussians of variance $\sigma_t^2$, starting from $t = 1$ with a simple complex Gaussian distribution of variance $\sigma^2_1 = \sigma^2_{\mathrm{max}}$ that can be sampled directly
until the lowest noise scale $\sigma_0=\sigma_{\mathrm{min}}$ is reached, where the learned score is assumed to approximate the true prior score well.

A straightforward idea is to formulate a diffusion process for the posterior in the same way as for the prior by adding Gaussian noise \cite{Kawar_Adv.NeuralInf.Process.Syst._2022}. To be able to use the learned prior, at each time $t$ the diffused posterior $p(\mathbf{ x}_t | \mathbf{y})$ should factorize into the diffused prior $p(\mathbf{ x}_t)$ and the conditional probability for measuring $\mathbf{y}$ when observing a perturbed sample $\mathbf{ x}_t$. By marginalization over the unknown noiseless images $\mathbf{x}$ one arrives at
\begin{align}
    p^{\mathrm{diffused}}(\mathbf y| \mathbf{ x}_t) = \int p(\mathbf y| \mathbf x) p(\mathbf x | \mathbf{ x}_t) d \mathbf{x}~.
    \label{eq:diff_lik}
\end{align}
We call this approach \textit{Diffused Posterior}. However, this likelihood term needs to be approximated in practice, as it makes use of the posterior for the denoising problem $p(\mathbf x | \mathbf{ x}_t)$, which is simpler than the full posterior but still intractable. This problem has led to various approximations for the term $p^{\mathrm{diffused}}(\mathbf y| \mathbf{ x}_t)$. Jalal et al. \cite{Jalal_NIPS_2021} introduced a weighting term $\kappa_t$ added to the data noise during the diffusion process:
\begin{align}
    p^{\mathrm{annealed}}(\mathbf y | \mathbf{ x}_t) \propto \exp\left(-\frac{1}{1 + \kappa_t}\lVert\mathbf{y}-A\mathbf{x}_t\rVert_2^2\right),
    \label{eq:diff_lik_annelead}
\end{align}
where $\kappa_t \to 0$ for decreasing noise scale $\sigma_t$ which we call \textit{Annealed Likelihood}. \citeauthor{Chung_Elev.Int.Conf.Learn.Represent._2023} \cite{Chung_Elev.Int.Conf.Learn.Represent._2023} used the score output by the network to calculate the expectation $\mathbb E[\mathbf{ x}_0|\mathbf{ x}_t]$ in order to approximate $p(\mathbf{x}|\mathbf{ x}_t) \approx \delta(\mathbf{x} - \mathbb E[
\mathbf{ x}_0|\mathbf{ x}_t])$ at every noise scale, i.e.
\begin{align}
    p^{\mathrm{DPS}}(\mathbf y | \mathbf{ x}_t) \propto \exp\left(-\lVert\mathbf{y}-A\mathbb E[\mathbf{ x}_0|\mathbf{ x}_t]\rVert_2^2\right).
    \label{eq:diff_lik_dps}
\end{align}

%The expectation at noise scale $\sigma_t$ is obtained by Tweedies formula $\mathbb E[\mathbf x_0|\mathbf x_k] = \mathbf x_k + \sigma_t^2 \nabla_\mathbf{x} \log p(\mathbf x_k)$ \cite{efron_tweedies_2011}.

While diffusing the posterior distribution is conceptually straightforward but practically challenging, an alternative approach is to use a different reverse process so long as the correct posterior distribution is obtained at $t=0$. This insight was first described in \citeauthor{Sohl-Dickstein_Proc.32ndInt.Conf.Mach.Learn._2015} \cite{Sohl-Dickstein_Proc.32ndInt.Conf.Mach.Learn._2015}, in which
the diffusion process was modified by multiplication with a function (e.g., the likelihood term). Hence, the learned prior distribution $p(\mathbf{ x}_t)$ can be multiplied by a
modified likelihood term $p(\mathbf{y}|\mathbf{ x}_t)$, which smoothly varies with $t$ and is equal to the true likelihood $p(\mathbf{y}|\mathbf{x})$ at $t=0$.
Here, we investigate the use of the original likelihood term in \eqref{eq:posterior}, i.e.
\begin{align}
    p^{\mathrm{exact}}(\mathbf y | \mathbf{ x}_t) \propto \exp\left(-\lVert\mathbf{y}-A\mathbf{x}_t\rVert_2^2\right),
    \label{eq:diff_lik_exact}
\end{align}
for the whole diffusion process, at every noise scale. Previously, this has been seen as a inacurrate approximation of the diffused likelihood in \eqref{eq:diff_lik}. However, here we use the original likelihood term from \eqref{eq:posterior} to construct a diffusion process not based on \eqref{eq:diff_lik} that, when understood as a different process, does not require any approximation. Hence, we call the likelihood \textit{Exact Likelihood}.

A brief discussion how this approach can be translated to the variance preserving formulation of diffusion models is provided in Supporting Information S1.
To motivate this choice, we show the effect of the choice of the likelihood for a 2D toy model in \figref{fig:toy_pdf_sampling}A that can be analytically computed similar to \cite{Gungor__2024}.

In contrast to the \textit{Annealed Likelihood} or \textit{Diffused Likelihood}, the \textit{Exact Likelihood} approach smoothes the posterior distribution only in directions that are not or weakly constrained by the measurement.
Importantly, all three likelihoods lead to the same, correct posterior distribution at the lowest noise level.

\begin{figure}
	\centering
	\includegraphics[width=\linewidth]{Figures/figure_01_analytic.pdf}
	\caption{
	A: Analytical diffusion process for a 2D toy model with a 1D linear measurement $A=(1\quad-1)^T$.
	The prior distribution is a mixture of 2D Gaussians forming a circle, which models the data manifold.
	Likelihood modifications corresponding to the diffused posterior and annealing are compared to the exact likelihood.
	All methods have the same posterior distribution at the minimum noise scale.
	B: Samples of the same 2D toy model with annealed (left) and exact (right) likelihood sampled with ULA for 10 and 500 iterations and pULA for 10 iterations.
	Samples are shown for noise levels $\sigma = [1, 0.2, 0.05, 0.01]$}
	\label{fig:toy_pdf_sampling}
\end{figure}

\subsection{Preconditioning of ULA}

A naive use of the \textit{Exact Likelihood} method is impractical because some directions are then constrained by the data already at high noise scales, implying a large Lipschitz constant corresponding to the maximum eigenvalue of $A^H A$ in the posterior score, exacerbated in non-Cartesian sampling. This then prevents the use of larger step sizes, leading to  slow convergence. To solve this issue, we propose a preconditioned ULA (pULA) with the update rule \cite{Roberts_MethodologyandComputinginAppliedProbability_2002,Corbineau_IEEESignalProcess.Lett._2019,Marnissi_IEEETrans.SignalProcess._2020,Bhattacharya_arXiv_2024}:
\begin{align}
    \bvec{x}_t^{k+1}=\bvec{x}_t^k+\gamma{M}_{t}\left[A^H(\bvec{y} - A\bvec{x}_t^k) + \nabla_{\bar{\mathbf{x}}}\log p_{t}(\bvec{x}_t^k)\right]+\sqrt{2\gamma}\mathbf{z}^k&&\mathbf{z}^k\sim\mathbb{C}\mathcal{N}(\bvec{0},M_{t}^{-1})\,,
    \label{eq:pula}
\end{align}
with a preconditioning matrix $M_{t}$ specifically adapted to linear inverse problems, i.e. ${M}_{t}=(A^HA+\sigma_t^{-2}\mathbb{I})^{-1}$.
Here, the superscript denotes the $k$th Langevin update per diffusion time $t$.
At $t=1$, the reverse-diffusion process is started with a sample from the posterior distribution corresponding to a flat Gaussian prior with variance $\sigma_1^2 = \sigma_{\mathrm{max}}^2$, i.e.
\begin{align}
    \bvec{x}_{t=1}^{k=0} \sim \mathbb{C}\mathcal{N}({M}_{1} \, {A}^H\bvec{y}, M_{1}^{-1})\,.
\end{align}
At each noise level $\sigma_t$, we perform $K$ Langevin steps of pULA before the noise level is reduced, and pULA is initialized with the last sample from the higher noise level.
%This corresponds to the predictor-corrector approach of \citeauthor{Song_ICLR_2021} \cite{Song_ICLR_2021} without a predictor step.

Our choice of the precondition matrix $M_{t}$ is based on the following observation:
The preconditioner should approximate the inverse Hessian of the negative log-posterior density.
The Hessian of the negative exact log-likelihood is given by $A^H A$ and we approximated the Hessian of the learned negative log prior by $\sigma_t^{-2}\mathbb{I}$, which corresponds to a Gaussian prior with variance $\sigma_t^2$.
In combination with the likelihood, the preconditioner allows larger updates
in directions of small singular values of $A$, i.e., those which are only weakly determined by the measurements. In these directions, the preconditioner effectively yields a step size proportional to the variance of the diffusion noise $\sigma_t^2$ as empirically found to work well for prior sampling \cite{Song_Adv.NeuralInf.Process.Syst._2019}.

The initialization and update rule of pULA require drawing samples $\mathbf{z}$ from the distribution $\mathbb{C}\mathcal{N}(\bvec{0}, {M}_{t}^{-1})$. Those can be drawn efficiently using \cite{Papandreou_Adv.NeuralInf.Process.Syst._2010}
\begin{align}
\bvec{z}
&= {M}_{t}\left({A}^H\bvec{n}_1 + \sigma_t^{-1}\bvec{n}_2\right) \quad
\begin{pmatrix}
\bvec{n}_1 \\ \bvec{n}_2
\end{pmatrix} \sim \mathbb{C}\mathcal{N}(\bvec{0}, \mathbb{I})\,,
\label{eq:sampling}
\end{align}
similar to the pseudo replica method \cite{Robbins_Proc.ThirdBerkeleySymp.Math.Stat.Probab._1954}. For the initialization, the mean can be added to the zero-mean sample.
Inserting Equation \ref{eq:sampling} into Equation \ref{eq:pula}, the update of pULA for the posterior score can be written as
\begin{align}
\bvec{x}_t^{k+1} = \bvec{x}_t^k + \gamma {M}_{t}&
\left[
	A^H\left(\bvec{y} + \sqrt{\frac{2}{\gamma}}\bvec{n}^k_1 - A\bvec{x}_t^k\right)\right. \notag \\ &\left. + \nabla_{\bar{\mathbf{x}}} \log p(\bvec{x}_t^k) + \sqrt{\frac{2}{\gamma\sigma^2_t}}\bvec{n}^k_2
\right]\,,
\quad
\begin{pmatrix}
\bvec{n}^k_1 \\ \bvec{n}^k_2
\end{pmatrix} \sim \mathbb{C}\mathcal{N}(\bvec{0}, \mathbb{I})\,.
\end{align}
In this form, it is apparent that the pULA update requires only one application of $M_{t}$, which can be performed using the conjugate gradient (CG) method
without explicitly forming the preconditioning matrix.
 
\section{Methods}
\label{sec:methods}
\subsection{Data Processing and Network Training}
% training parameter

%\subsection{Data Processing}

For training, we used images reconstructed from the fully sampled k-space data of the multi-coil train folder of the fastMRI brain dataset \cite{Knoll_Radiol.Artif.Intell._2020} that have a $320\times 320$ matrix size.
To be consistent with the inference pipeline, the k-space data were pre-whitened using a noise covariance matrix estimate from the background patches of the fully sampled coil images. The data were then compressed to 12 virtual coils, and coil sensitivities were estimated with a low-resolution version of NLINV \cite{Uecker_Magn.Reson.Med._2008}. Using the corresponding reconstruction, a scaling for the k-space was estimated such that the final reconstruction is normalized to have a 99 percentile pixel magnitude of one.
A foreground mask for the images was estimated using ESPIRiT \cite{Uecker_Magn.Reson.Med._2014}.
Data for which this mask extends to the image boundary in the phase encoding direction were assumed to be corrupted, for example by motion, and removed.
After this step, the training dataset contains around 32,000 of the original 36,000 samples.
To suppress background noise, the final images for training were then reconstructed using FISTA with a small $\ell_1$-Wavelet regularization.
Using these high quality images, we trained a conditional denoising U-Net with conditional residual blocks \cite{Song_ICLR_2021} using Adam ($\text{learning rate} = 0.0001$, $\text{batch size} = 32$ and $\text{epochs} = 100$) implemented in BART \cite{Blumenthal_Magn.Reson.Med._2023}.
The sigma schedule was an exponential decay from $\sigma_{\mathrm{max}} = 100$ to $\sigma_{\mathrm{min}} = 0.01$.
From the denoising U-Net, the score network is obtained using Tweedie's formula \cite{Robbins_Proc.ThirdBerkeleySymp.Math.Stat.Probab._1954,Efron_J.Am.Stat.Assoc._2011}.

\subsection{MRI Data Acquisition}

MRI data of a healthy volunteer was acquired on a 3T scanner (Magnetom Vida, Siemens Healthineers, Erlangen, Germany)
after obtaining written informed consent and with approval of the local ethics committee.
We acquired fully sampled multi-slice Cartesian T2-weighted ($\text{TR}=\SI{6000}{\milli\second}$, $\text{TE}=\SI{98}{\milli\second}$, $\Delta\text{TE}=\SI{9.82}{\milli\second}$, $\text{FA} = 150^\circ$, $\text{ETL} = 16$, $\text{BW}=\SI{223}{\hertz\per\pixel}$) and T1-weighted ($\text{TR}=\SI{600}{\milli\second}$, $\text{TE}=\SI{6.70}{\milli\second}$, $\Delta\text{TE}=\SI{6.69}{\milli\second}$, $\text{FA} = 140^\circ$, $\text{ETL} = 2$, $\text{BW}=\SI{391}{\hertz\per\pixel}$) brain data with a 20-channel head coil using a Turbo Spin Echo sequence.

In addition, a radial T1-weighted scan was acquired using a stack-of-stars FLASH sequence with a RAGA \cite{Scholand_Magn.Reson.Med._2025} sampling scheme ($\text{TR}=\SI{8.0}{\milli\second}$, $\text{TE}=\SI{3.32}{\milli\second}$, $\text{FA} = \SI{10}{\degree}$, voxel size: $0.8\times 0.8\times \SI{3}{\cubic\milli\meter}$, $\text{BW}=\SI{780}{\hertz\per\pixel}$). Two repetitions with $987$ spokes each were acquired, \textit{i.e.}, two RAGA full frames. All datasets had a FoV of $\SI{250}{\milli\meter}$ and matrix size of $320\times 320$ pixels.

\subsection{Numerical Experiments}

We first tested the exact likelihood approach for a real-valued 2D toy model with a linear measurement operator $A=\begin{pmatrix}1&-1\end{pmatrix}^T$ and an analytical prior defined by a Gaussian mixture model. pULA with the exact likelihood was compared to ULA with exact and annealed likelihood, where in total $N=101$ noise scales were used with $\sigma_{\mathrm{max}} = 1$ and $\sigma_{\mathrm{min}} = 0.01$.

Then, the proposed exact likelihood method was compared to an $\ell_1$-Wavelet regularized parallel imaging reconstruction, DPS and the annealed likelihood approach on a slice of the T2-weighted dataset. For this,
the fully sampled k-space data were pre-whitened based on noise extracted from a background patch of the fully sampled coil images. The data were then retrospectively undersampled with both equispaced and randomized undersampling masks (acceleration 4 and 12) with a 16-line auto-calibration region. Similar to the processing of the training data, the undersampled data were coil compressed to 12 virtual channels, coils were estimated with NLINV, a foreground mask was computed with ESPIRiT, and a normalization scale was computed from the low-resolution NLINV reconstruction. Instead of scaling the k-space data, we absorb this normalization constant in the coil sensitivities, such that the k-space data stays white with unit variance noise and the scaling of the corresponding image remains consistent with normalization used for the training images.

The sampling techniques were implemented in BART. For the annealed likelihood, we choose the annealing parameter $\kappa_t$ such that
\begin{align}
    \frac{1}{1+\kappa_t} = \left(\frac{\sigma_{\mathrm{max}}^{-2}}{\lambda_{\mathrm{max}}(A^HA)}\right)^t\,,
\end{align}
where $\lambda_{\mathrm{max}}(A^HA)$ is the maximum eigenvalue of $A^HA$, which we estimate using power iterations.
At $t=1$, this choice balances the Lipschitz constants of the likelihood and the prior scores. For ULA of the annealed likelihood (aULA),
we chose the step size $\gamma = \gamma_{\mathrm{base}} \left[(1 + \kappa_t)^{-1}\lambda_{\mathrm{max}}(A^HA) + \sigma_t^{-2}\right]^{-1}$ with a base scaling $\gamma_{\mathrm{base}}=0.5$ in all experiments. For pULA, we use the same step size for all noise levels, namely, $\gamma = 0.5$. The implementation of DPS is detailed in Supporting Information Text S3.

For the diffusion reconstruction of the T2-weighted dataset, we exponentially reduced the diffusion noise level $\sigma_t$ from $\sigma_{\mathrm{max}}=10$ to $\sigma_{\mathrm{min}}=0.01$.
Ten samples were drawn and averaged using aULA, DPS and pULA with exact likelihood. $N=60$ noise levels were used for aULA and pULA with $K=8$ (aULA) and $K=4$ (pULA) Langevin iterations per noise level.
The preconditioner was applied with $N_{CG}=10$ CG iterations using a warmstarting strategy detailed in Supporting Information Text S2. For DPS, we used the predictor sampling \cite{Song_ICLR_2021} with the same amount of network evaluations compared to the exact likelihood approach. The weighting of the DPS likelihood was set to $\zeta'=0.2$, which yields best results for 12x acceleration.
If not stated otherwise, the same parameters were used for all other experiments.
In addition, we tested all methods starting at $\sigma_{\mathrm{max}}=1$ and $\sigma_{\mathrm{max}}=0.1$ and correspondingly reduced the number of noise levels to $N=40$ and $N=20$, respectively.
The regularization parameter for the undersampled $\ell_1$-Wavelet reconstruction was chosen by a grid search to obtain the best PSNR with respect to the fully sampled reconstruction. For all reconstructions, the time per sample was measured.
All images were normalized by multiplication with the RSS of the coils to obtain an RSS scaling of the final reconstructions.
Afterward, PSNR and SSIM values and error maps were computed with magnitude images relative to the fully sampled $\ell_1$-Wavelet reconstruction after multiplying all images with the foreground mask.

To investigate the robustness of the method under substantially different experimental conditions, we performed three additional experiments.
First, the number of virtual coils used for sampling/reconstruction of the T1-weighted dataset was decreased to 4 or 1 after the pre-processing described above, i.e., the full pre-processing was performed using 12 virtual coils.
Second, we performed sampling of the T2-weighted dataset with three-fold increased noise level. For DPS a grid search was conducted to determine the likelihood weighting $\zeta'$ with optimal PSNR.
Third, we performed reconstructions of the radial data, where we reduced the number of radial spokes from 987 spokes to 98, 49, and 24 spokes. To discard data during the transition towards the FLASH steady state, we used the data of the second frame only. The radial data was processed similar to the Cartesian data, but first an inverse FFT in partition direction was performed to allow for slice-wise processing. The gradient delays were corrected with RING \cite{Rosenzweig_Magn.Reson.Med._2019}. To compute the ESPIRiT foreground mask, k-space data was gridded first, whereas coil sensitivities were directly estimated from the radial data with NLINV.
In all cases, an $\ell_1$-Wavelet reconstruction was also performed.

All numerical experiments have been performed on a system with an AMD EPYC 9334 CPU and an Nvidia H100 GPU (80 GB HBM3, SXM). We record and compare total runtime for all algorithms (c.f. Supporting Information Text S4).
 
\section{Results}
\label{sed:results}

\begin{figure}
	\centering
	\includegraphics[width=\linewidth]{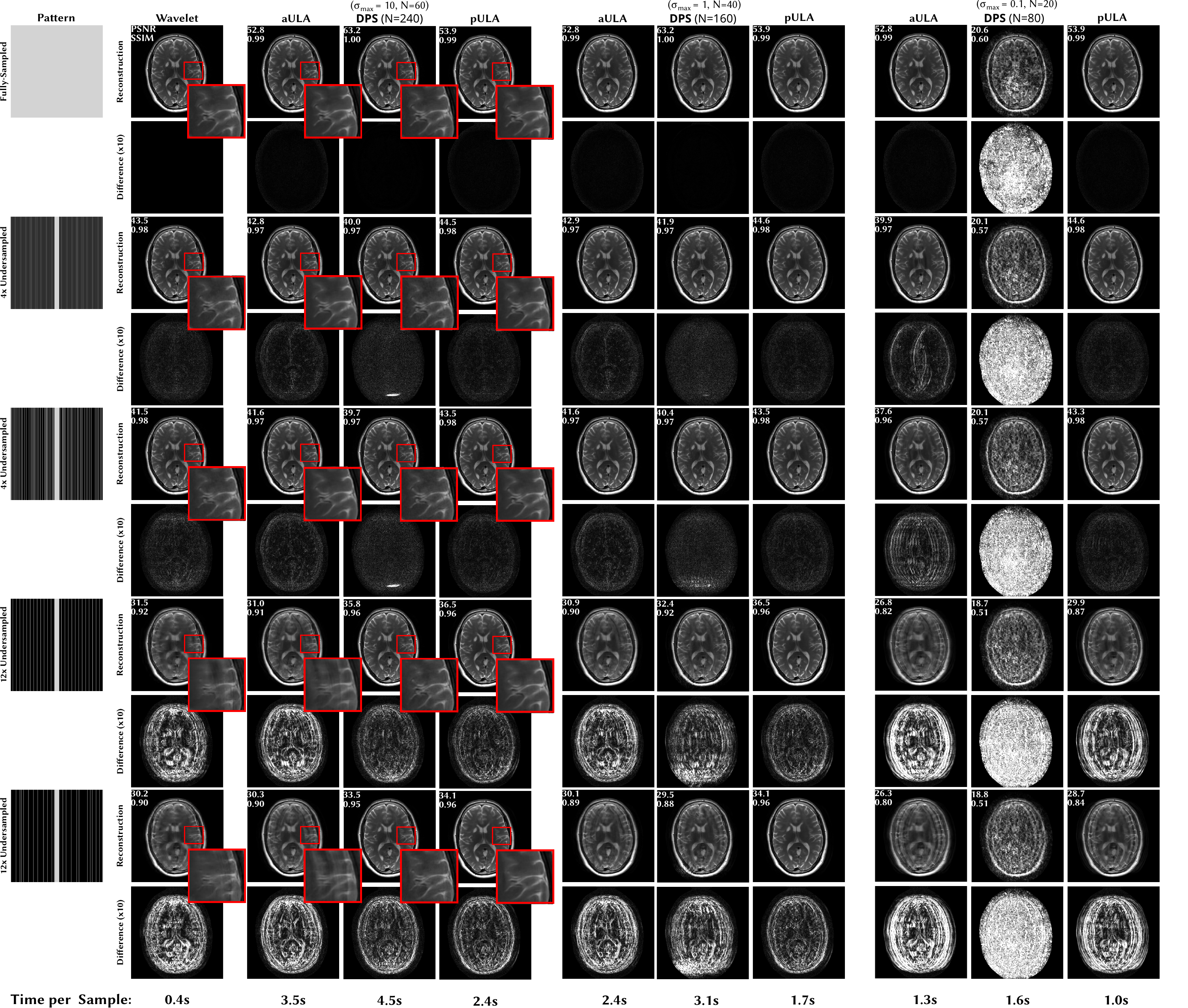}
	\caption{Reconstructions of a T2-weighted brain image for different undersampling patterns using $\ell_1$-Wavelet regularization and diffusion posterior sampling with annealed (pULA) and exact (ULA) likelihood. For $\ell_1$-Wavelet and $\sigma_{\max} = 10$ reconstruction a zoomed-in view for a high-detail region (highlighted in a red box) is shown. Error maps and PSNR/SSIM values are computed relative to the fully-sampled $\ell_1$-Wavelet reconstruction. Per noise level, $K=8$ ULA or $K=4$ pULA iterations were performed.}
	\label{fig:comp}
\end{figure}

\begin{figure}
	\centering
	\includegraphics[width=\linewidth]{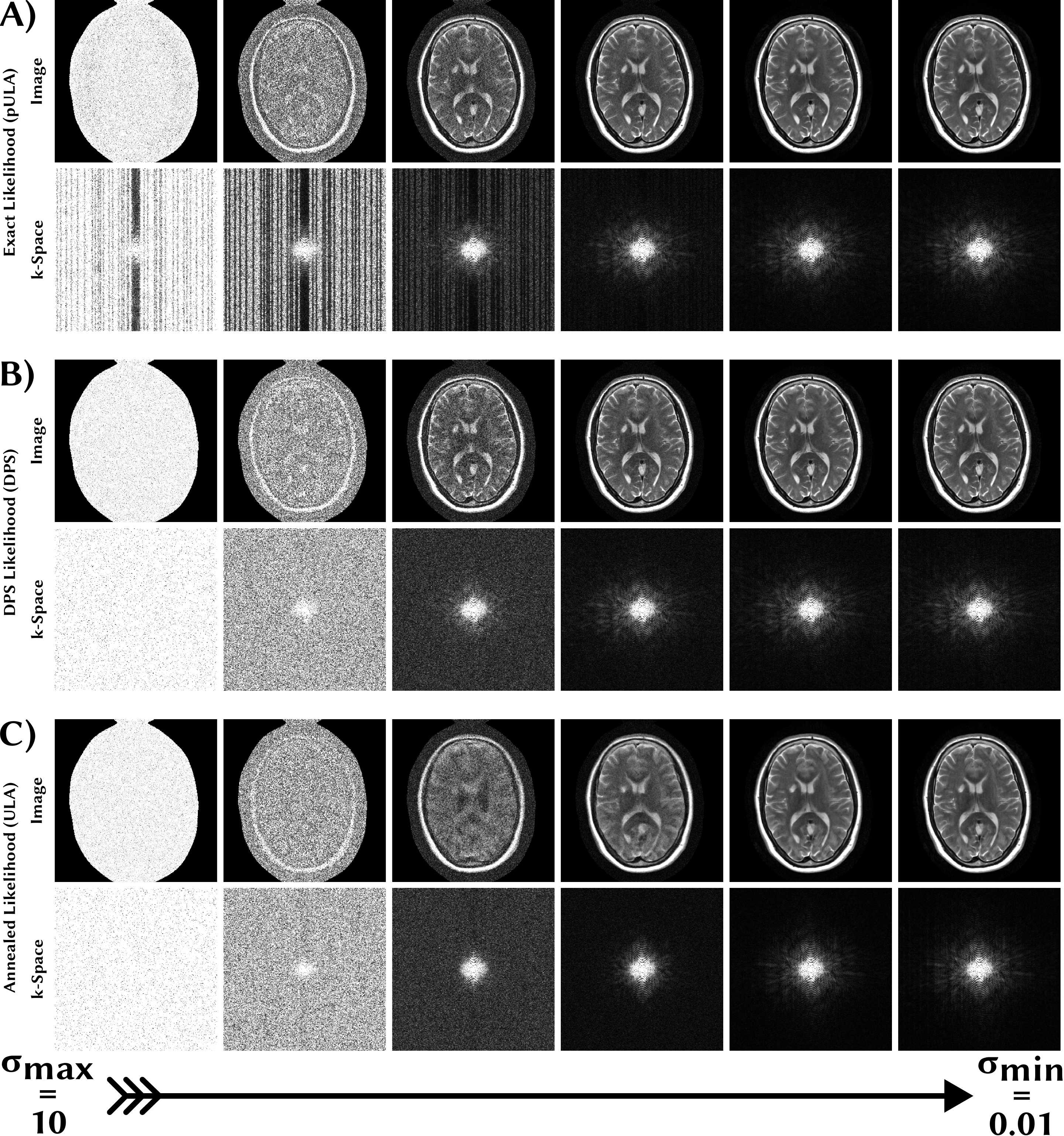}
	\caption{Reverse diffusion process with exact (A), DPS (B) and annealed (C) likelihood for a T2-weighted brain image sampled from 8x random undersampled data.}
	\label{fig:diffusion_process}
\end{figure}

\begin{figure}
	\centering
	\includegraphics[width=\linewidth]{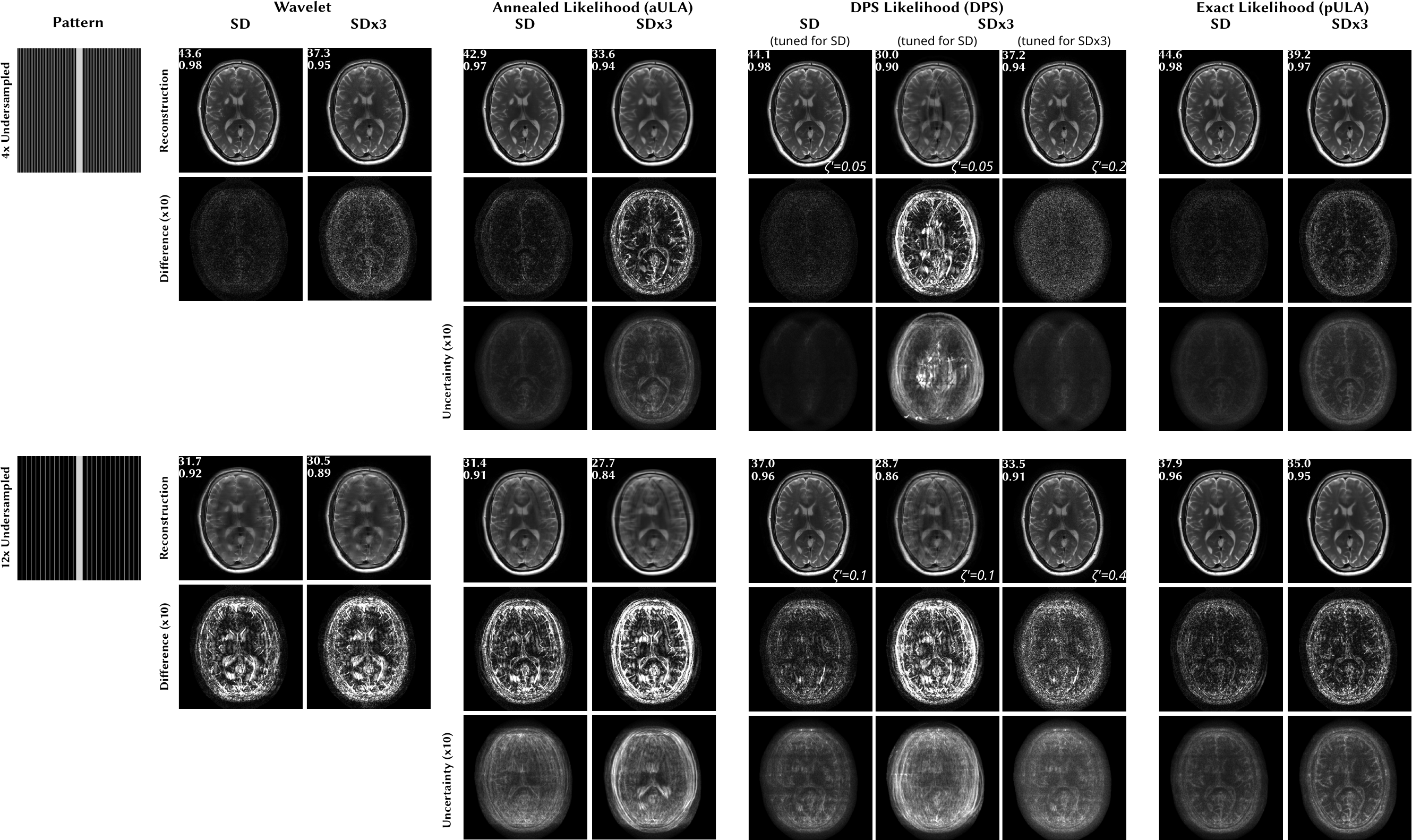}
	\caption{Comparison of  reconstructions from original k-space data (SD) and three-fold increased noise level (SDx3). The weighting of the DPS likelihood ($\zeta'$) was tuned for each noise level. Using $\zeta'$ tuned for the wrong noise level shows a significant drop in performance. aULA and pULA adapt to the noise level without the need for tuning, and pULA yields the best results in PSNR for both noise levels.}
	\label{fig:noise_comp}
\end{figure}

\begin{figure}
	\centering
	\includegraphics[width=\linewidth]{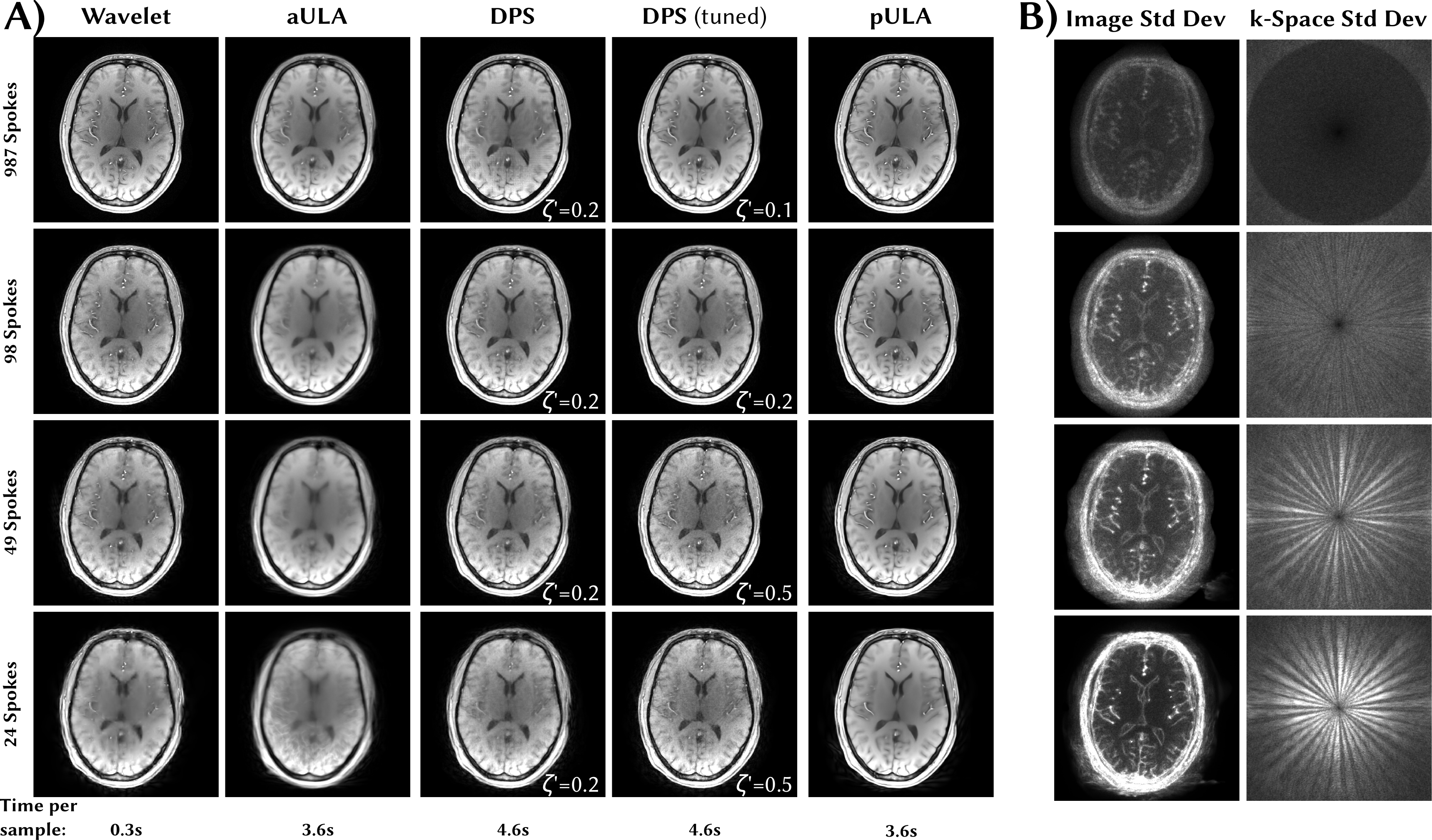}
	\caption{Brain image reconstructed from radially acquired FLASH data using different undersampling factors. A: Reconstruction with $\ell_1$-Wavelet regularization, the average of ten samples from the posterior drawn with annealed likelihood (aULA), DPS, and the exact likelihood (pULA). DPS was once performed with the weighting $\zeta'=0.2$ from the Cartesian experiments and once with $\zeta'$ tuned per undersampling. B:  Pixel-wise standard deviation map of the exact likelihood approach in image space and k-space.}
	\label{fig:noncart}
\end{figure}

Figure \ref{fig:toy_pdf_sampling}B shows
the 2D toy example of the sampling results for the reverse diffusion process comparing the annealed and the exact likelihood method for (p)ULA. In total 1000 samples were drawn. The annealed posterior adapts during the diffusion process, whereas the exact posterior does not change. For time zero, all methods converge to the correct posterior distribution. However, there are different convergence speeds. For the exact likelihood, ULA needs a small step size and therefore more iterations ($K = 500$) to converge to the correct distribution. pULA with exact likelihood shows faster convergence, similar to the annealed approach ($K = 10$).

Sampling results for the Cartesian data are shown in Figure~\ref{fig:comp}.
While all methods yield good reconstructions, the exact likelihood with pULA
requires overall less computation time (c.f. Supporting Information Text S4 for
the extended analysis) and is consistently outperforming the
annealed likelihood and DPS method in terms of PSNR and SSIM as well as showing
reduced visible differences in the error maps. The exact likelihood approach is faster,
as it requires less network evaluations compared to aULA and no backpropagation through the network as DPS does.
Moreover, the fast convergence of the exact likelihood approach allows for starting
at smaller noise scales ($\sigma_{\mathrm{max}}=0.1$ for 4x acceleration and
$\sigma_{\mathrm{max}}=1$ for 12x acceleration), which further reduces the reconstruction
time compared to the other methods, which require starting at higher noise.

The diffusion processes for $8 \times$ randomly undersampled k-space data are shown in Figure~\ref{fig:diffusion_process}.
When using the exact likelihood, even at high noise levels, noise is only visible in k-space
locations that are not acquired. In contrast, for the annealed likelihood and DPS, noise
appears in all k-space lines at the beginning of the reverse diffusion process. At a high noise
scale, the annealed method first generates what seems to resemble a T1-weighted image with dark
CSF, before the data term becomes dominant enough to guide the reconstruction towards the correct
image. This behavior does not occur when using the exact likelihood.

The performance of all methods with noise-corrupted k-space data is presented in Figure~\ref{fig:noise_comp}.
The exact likelihood approach combined with pULA delivers consistently strong results across all scenarios,
including high undersampling patterns and different noise corruption. Notably, DPS requires tuning of the likelihood
weighting $\zeta'$ for each noise level to achieve good performance (c.f. Supporting Figure S4).
Reconstructions for different numbers of virtual coils are shown in Supporting Figure S5.

The results of the non-Cartesian reconstruction are shown in Figure \ref{fig:noncart}.
Reconstructions with the annealed likelihood are blurred due to the ill-conditioning of the non-Cartesian SENSE model, even when all
spokes are used. The DPS approach shows comparable results to the exact likelihood approach, however, only for specific weightings
$\zeta'$ tuned for each undersampling factor (c.f. Supporting Figure S6).
The pixel-wise standard deviation shows high uncertainty in regions containing small vessels.

%of the pULA approach are rich in details even in the 40-fold undersampled case, while the average is slightly blurred. The pixel-wise standard deviation shows high uncertainty in regions containing small vessels.

%, which is expected, as due to the high noise (visible in the $\ell_1$-Wavelet reconstruction).
%many images can explain the data.

\section{Discussion}
\label{sec:discussion}

In this work, we addressed the problem that existing methods for posterior sampling
for MRI reconstruction are rather slow and necessitate cumbersome parameter tuning
to balance computation time and errors depending on the ill-posedness of the inverse
problem. We investigated the use of an exact likelihood term rather than adjusting
the likelihood in the diffusion process as done in previous publications.

In a 2D toy example, one can see that a naive use of the exact likelihood leads
to slow convergence because the data restricts the problem space along one dimension,
preventing large step sizes. While the annealed likelihood avoids this problem as
it reduces the maximum eigenvalue, it comes at the cost of not utilizing the data
properly at later diffusion times and requires careful parameter tuning. As we
showed, slow convergence can be addressed effectively by preconditioning. The exact
likelihood with pULA then shows similar performance in the 2D toy example as the annealed
likelihood method. This observation can  be extended to real MRI data where sampling
with exact likelihood and pULA then consistently outperforms the annealing and DPS, and shows robust behavior for different undersampling masks,
numbers of virtual coils, and non-Cartesian radial MRI while using the same step
size, number of iterations, and the same maximum noise level in all scenarios.

Similar to other posterior sampling methods, the fundamental assumption of the exact likelihood
is that the MRI measurement is accurately described by the SENSE forward model and additive
Gaussian noise. In case of a model mismatch, e.g. due to motion or spike noise, it may be
beneficial to adapt the forward model to include motion \cite{Levac_Magn.Reson.Med._2024}
or to adaptively reweight the data consistency term to respect potential outliers
in case of spike noise \cite{Hen__2025}.

As our results show, the exact likelihood approach with pULA is significantly faster
than the annealed or DPS approach, while achieving the same or superior reconstruction
quality. This is because fewer expensive network evaluations are required for
convergence. Both, network evaluations and
CG steps scale (log)-linearly with the problem size, such that we expect similar
benefits of the exact likelihood approach with pULA for 3D reconstructions. However,
poorer problem conditioning for 3D-non-Cartesian sampling could require more CG
iterations for preconditioning.

Using the exact likelihood naturally restricts sampling to the problem space determined by
the data and eliminates the need for likelihood approximation or hyperparameter tuning.
Although this formulation requires small step sizes in practice, this limitation can be
effectively mitigated through preconditioning. In this sense, the exact likelihood approach
combined with pULA constitutes a unified method that enables efficient and principled sampling
from the posterior distribution, while automatically adapting to different acquisition
settings and measurement noise levels without requiring hyperparameter tuning.

\section{Conclusion}
\label{sec:conclusion}

The proposed exact likelihood with preconditioned ULA enables fast and robust posterior
sampling for different MRI reconstruction problems without parameter tuning. Especially
ill-conditioned problems such as radial MRI benefit from the increased convergence speed.

%TC:ignore

\section*{Conflict of Interest}
The authors declare no competing interests.

\section*{Data Availability Statement}
In the spirit of reproducible research, the code to reproduce the results of this paper is available at \url{https://gitlab.tugraz.at/ibi/mrirecon/papers/dps-pula} (Version v0.2). All reconstructions have been performed with BART (v1.0.00), available at \url{https://github.com/mrirecon/bart}.
The data used in this study is available at Zenodo (DOI: \doi{10.5281/zenodo.17739731}).

\section*{Acknowledgements}

\printfunding

\printbibliography

@online{Bhattacharya_arXiv_2024,
  title = {Fast {{Sampling}} and {{Inference}} via {{Preconditioned Langevin Dynamics}}},
  author = {Bhattacharya, Riddhiman and Jiang, Tiefeng},
  date = {2024-03-29},
  eprint = {2310.07542},
  eprinttype = {arXiv},
  eprintclass = {stat},
  doi = {10.48550/arXiv.2310.07542},
  url = {http://arxiv.org/abs/2310.07542},
  urldate = {2025-10-16},
  pubstate = {prepublished}
}

@article{Blumenthal_Magn.Reson.Med._2023,
  title = {Deep, Deep Learning with {{BART}}},
  author = {Blumenthal, Moritz and Luo, Guanxiong and Schilling, Martin and Holme, H. Christian M. and Uecker, Martin},
  date = {2023},
  journaltitle = {Magn. Reson. Med.},
  volume = {89},
  number = {2},
  eprint = {https://onlinelibrary.wiley.com/doi/pdf/10.1002/mrm.29485},
  pages = {678--693},
  doi = {10.1002/mrm.29485},
  url = {https://onlinelibrary.wiley.com/doi/abs/10.1002/mrm.29485},
  langid = {english}
}

@online{Chung__2025,
  title = {Diffusion Models for Inverse Problems},
  author = {Chung, Hyungjin and Kim, Jeongsol and Ye, Jong Chul},
  date = {2025-08-04},
  eprint = {2508.01975},
  eprinttype = {arXiv},
  eprintclass = {cs},
  doi = {10.48550/arXiv.2508.01975},
  url = {http://arxiv.org/abs/2508.01975},
  urldate = {2025-12-04},
  pubstate = {prepublished}
}

@inproceedings{Chung_Elev.Int.Conf.Learn.Represent._2023,
  title = {Diffusion {{Posterior Sampling}} for {{General Noisy Inverse Problems}}},
  booktitle = {{{ICLR}} 2023: {{The Eleventh International Conference}} on {{Learning Representations}}},
  author = {Chung, Hyungjin and Kim, Jeongsol and Mccann, Michael T and Klasky, Marc L and Ye, Jong Chul},
  date = {2023},
  volume = {11},
  eventtitle = {International {{Conference}} on {{Learning Representations}}},
  langid = {english}
}

@inproceedings{Chung_ICLR_2024,
  title = {Decomposed {{Diffusion Sampler}} for {{Accelerating Large-Scale Inverse Problems}}},
  author = {Chung, Hyungjin and Lee, Suhyeon and Ye, Jong Chul},
  date = {2024},
  url = {https://openreview.net/forum?id=DsEhqQtfAG},
  urldate = {2026-01-20},
  eventtitle = {The {{Twelfth International Conference}} on {{Learning Representations}}},
  langid = {english}
}

@article{Chung_Med.ImageAnal._2022,
  title = {Score-Based Diffusion Models for Accelerated {{MRI}}},
  author = {Chung, Hyungjin and Ye, Jong Chul},
  date = {2022-08},
  journaltitle = {Med Image Anal},
  volume = {80},
  pages = {102479},
  issn = {1361-8415},
  doi = {10.1016/j.media.2022.102479},
  url = {https://www.sciencedirect.com/science/article/pii/S1361841522001268},
  urldate = {2024-03-23},
  langid = {english}
}

@article{Corbineau_IEEESignalProcess.Lett._2019,
  title = {Preconditioned {{P-ULA}} for {{Joint Deconvolution-Segmentation}} of {{Ultrasound Images}}},
  author = {Corbineau, Marie-Caroline and Kouam\'e, Denis and Chouzenoux, Emilie and Tourneret, Jean-Yves and Pesquet, Jean-Christophe},
  date = {2019-10},
  journaltitle = {IEEE Signal Processing Letters},
  volume = {26},
  number = {10},
  pages = {1456--1460},
  issn = {1558-2361},
  doi = {10.1109/LSP.2019.2935610},
  url = {https://ieeexplore.ieee.org/document/8801938/similar},
  urldate = {2025-10-16}
}

@article{Dalalyan_J.R.Stat.Soc.Ser.BStat.Methodol._2017,
  title = {Theoretical {{Guarantees}} for {{Approximate Sampling}} from {{Smooth}} and {{Log-Concave Densities}}},
  author = {Dalalyan, Arnak S.},
  date = {2017-06-01},
  journaltitle = {Journal of the Royal Statistical Society Series B: Statistical Methodology},
  volume = {79},
  number = {3},
  pages = {651--676},
  issn = {1369-7412, 1467-9868},
  doi = {10.1111/rssb.12183},
  url = {https://academic.oup.com/jrsssb/article/79/3/651/7040689},
  urldate = {2025-11-22},
  langid = {english}
}

@online{Daras_arxiv_2024,
  title = {A Survey on Diffusion Models for Inverse Problems},
  author = {Daras, Giannis and Chung, Hyungjin and Lai, Chieh-Hsin and Mitsufuji, Yuki and Ye, Jong Chul and Milanfar, Peyman and Dimakis, Alexandros G. and Delbracio, Mauricio},
  date = {2024-09-30},
  eprint = {2410.00083},
  eprinttype = {arXiv},
  eprintclass = {cs},
  doi = {10.48550/arXiv.2410.00083},
  url = {http://arxiv.org/abs/2410.00083},
  urldate = {2025-10-16},
  pubstate = {prepublished},
  version = {1}
}

@article{Efron_J.Am.Stat.Assoc._2011,
  title = {Tweedie's {{Formula}} and {{Selection Bias}}},
  author = {Efron, Bradley},
  date = {2011-12},
  journaltitle = {J. Am. Stat. Assoc.},
  volume = {106},
  number = {496},
  pages = {1602--1614},
  issn = {0162-1459, 1537-274X},
  doi = {10.1198/jasa.2011.tm11181},
  url = {http://www.tandfonline.com/doi/abs/10.1198/jasa.2011.tm11181},
  urldate = {2025-11-12},
  langid = {english}
}

@online{Gungor__2024,
  title = {Bayesian {{Conditioned Diffusion Models}} for {{Inverse Problems}}},
  author = {G\"ung\"or, Alper and Bilecen, Bahri Batuhan and \c Cukur, Tolga},
  date = {2024-06-14},
  eprint = {2406.09768},
  eprinttype = {arXiv},
  eprintclass = {cs},
  doi = {10.48550/arXiv.2406.09768},
  url = {http://arxiv.org/abs/2406.09768},
  urldate = {2026-01-20},
  pubstate = {prepublished}
}

@online{Hen__2025,
  title = {Robust {{Posterior Diffusion-based Sampling}} via {{Adaptive Guidance Scale}}},
  author = {Hen, Liav and Tirer, Tom and Giryes, Raja and Abu-Hussein, Shady},
  date = {2025-11-23},
  eprint = {2511.18471},
  eprinttype = {arXiv},
  eprintclass = {cs},
  doi = {10.48550/arXiv.2511.18471},
  url = {http://arxiv.org/abs/2511.18471},
  urldate = {2026-03-03},
  pubstate = {prepublished}
}

@inproceedings{Ho_NIPS_2020,
  title = {Denoising Diffusion Probabilistic Models},
  booktitle = {Advances in Neural Information Processing Systems},
  author = {Ho, Jonathan and Jain, Ajay and Abbeel, Pieter},
  editor = {Larochelle, H. and Ranzato, M. and Hadsell, R. and Balcan, M.F. and Lin, H.},
  date = {2020},
  volume = {33},
  pages = {6840--6851},
  publisher = {Curran Associates, Inc.},
  url = {https://proceedings.neurips.cc/paper_files/paper/2020/file/4c5bcfec8584af0d967f1ab10179ca4b-Paper.pdf}
}

@inproceedings{Holliber_Proc.Annu.Meet.ISMRM_2025,
  title = {Unadjusted {{Langevin Sampling}} for {{Uncertainty Estimation}} in {{MRI Reconstruction}} - {{Theory}} and {{Numerical Validation}}},
  booktitle = {Proceedings of the {{Annual Meeting}} of {{ISMRM}}},
  author = {Holliber, Tina and Blumenthal, Moritz and Uecker, Martin},
  date = {2025},
  pages = {2603}
}

@inproceedings{Jalal_NIPS_2021,
  title = {Robust Compressed Sensing {{MRI}} with Deep Generative Priors},
  booktitle = {Advances in Neural Information Processing Systems},
  author = {Jalal, Ajil and Arvinte, Marius and Daras, Giannis and Price, Eric and Dimakis, Alexandros G and Tamir, Jon},
  editor = {Ranzato, M. and Beygelzimer, A. and Dauphin, Y. and Liang, P.S. and Vaughan, J. Wortman},
  date = {2021},
  volume = {34},
  pages = {14938--14954},
  publisher = {Curran Associates, Inc.},
  url = {https://proceedings.neurips.cc/paper/2021/file/7d6044e95a16761171b130dcb476a43e-Paper.pdf}
}

@article{Janati_Phil.Trans.R.Soc.A._2025,
  title = {Bridging Diffusion Posterior Sampling and {{Monte Carlo}} Methods: A Survey},
  shorttitle = {Bridging Diffusion Posterior Sampling and {{Monte Carlo}} Methods},
  author = {Janati, Yazid and Moulines, Eric and Olsson, Jimmy and Oliviero-Durmus, Alain},
  date = {2025-06-19},
  journaltitle = {Phil. Trans. R. Soc. A.},
  volume = {383},
  number = {2299},
  pages = {20240331},
  issn = {1364-503X, 1471-2962},
  doi = {10.1098/rsta.2024.0331},
  url = {https://royalsocietypublishing.org/doi/10.1098/rsta.2024.0331},
  urldate = {2025-10-15},
  langid = {english}
}

@inproceedings{Karras_Adv.NeuralInf.Process.Syst._2022,
  title = {Elucidating the {{Design Space}} of {{Diffusion-Based Generative Models}}},
  booktitle = {Advances in {{Neural Information Processing Systems}}},
  author = {Karras, Tero and Aittala, Miika and Aila, Timo and Laine, Samuli},
  date = {2022-12-06},
  volume = {35},
  pages = {26565--26577},
  url = {https://proceedings.neurips.cc/paper_files/paper/2022/hash/a98846e9d9cc01cfb87eb694d946ce6b-Abstract-Conference.html},
  urldate = {2025-12-04},
  langid = {english}
}

@inproceedings{Kawar_Adv.NeuralInf.Process.Syst._2022,
  title = {Denoising {{Diffusion Restoration Models}}},
  booktitle = {Advances in {{Neural Information Processing Systems}}},
  author = {Kawar, Bahjat and Elad, Michael and Ermon, Stefano and Song, Jiaming},
  date = {2022-12-06},
  volume = {35},
  pages = {23593--23606},
  url = {https://papers.nips.cc/paper_files/paper/2022/hash/95504595b6169131b6ed6cd72eb05616-Abstract-Conference.html},
  urldate = {2025-12-04},
  langid = {english}
}

@article{Knoll_Radiol.Artif.Intell._2020,
  title = {{{fastMRI}}: {{A Publicly Available Raw}} k-{{Space}} and {{DICOM Dataset}} of {{Knee}}                     {{Images}} for {{Accelerated MR Image Reconstruction Using Machine}}                     {{Learning}}},
  shorttitle = {{{fastMRI}}},
  author = {Knoll, Florian and Zbontar, Jure and Sriram, Anuroop and Muckley, Matthew J. and Bruno, Mary and Defazio, Aaron and Parente, Marc and Geras, Krzysztof J. and Katsnelson, Joe and Chandarana, Hersh and Zhang, Zizhao and Drozdzalv, Michal and Romero, Adriana and Rabbat, Michael and Vincent, Pascal and Pinkerton, James and Wang, Duo and Yakubova, Nafissa and Owens, Erich and Zitnick, C. Lawrence and Recht, Michael P. and Sodickson, Daniel K. and Lui, Yvonne W.},
  date = {2020-01},
  journaltitle = {Radiology: Artificial Intelligence},
  volume = {2},
  number = {1},
  pages = {e190007},
  publisher = {Radiological Society of North America},
  doi = {10.1148/ryai.2020190007},
  url = {https://pubs.rsna.org/doi/10.1148/ryai.2020190007},
  urldate = {2024-04-02}
}

@online{Kreutz-Delgado__2009,
  title = {The {{Complex Gradient Operator}} and the {{CR-Calculus}}},
  author = {Kreutz-Delgado, Ken},
  date = {2009},
  doi = {10.48550/ARXIV.0906.4835},
  url = {https://arxiv.org/abs/0906.4835},
  urldate = {2025-11-24},
  pubstate = {prepublished},
  version = {1}
}

@article{Levac_Magn.Reson.Med._2024,
  title = {Accelerated Motion Correction with Deep Generative Diffusion Models},
  author = {Levac, Brett and Kumar, Sidharth and Jalal, Ajil and Tamir, Jonathan I.},
  date = {2024},
  journaltitle = {Magn. Reson. Med.},
  volume = {92},
  number = {2},
  pages = {853--868},
  issn = {1522-2594},
  doi = {10.1002/mrm.30082},
  url = {https://onlinelibrary.wiley.com/doi/abs/10.1002/mrm.30082},
  urldate = {2026-03-03},
  langid = {english}
}

@article{Luo_Magn.Reson.Med._2023,
  title = {Bayesian {{MRI}} Reconstruction with Joint Uncertainty Estimation Using Diffusion Models},
  author = {Luo, Guanxiong and Blumenthal, Moritz and Heide, Martin and Uecker, Martin},
  date = {2023},
  journaltitle = {Magn. Reson. Med.},
  volume = {90},
  number = {1},
  pages = {295--311},
  doi = {10.1002/mrm.29624}
}

@article{Marnissi_IEEETrans.SignalProcess._2020,
  title = {Majorize--{{Minimize Adapted Metropolis}}--{{Hastings Algorithm}}},
  author = {Marnissi, Yosra and Chouzenoux, Emilie and Benazza-Benyahia, Amel and Pesquet, Jean-Christophe},
  date = {2020},
  journaltitle = {IEEE Trans. Signal Process.},
  volume = {68},
  pages = {2356--2369},
  issn = {1053-587X, 1941-0476},
  doi = {10.1109/TSP.2020.2983150},
  url = {https://ieeexplore.ieee.org/document/9050537/},
  urldate = {2025-11-23}
}

@inproceedings{Papandreou_Adv.NeuralInf.Process.Syst._2010,
  title = {Gaussian Sampling by Local Perturbations},
  booktitle = {Advances in Neural Information Processing Systems},
  author = {Papandreou, George and Yuille, Alan L},
  editor = {Lafferty, J. and Williams, C. and Shawe-Taylor, J. and Zemel, R. and Culotta, A.},
  date = {2010},
  volume = {23},
  publisher = {Curran Associates, Inc.},
  url = {https://proceedings.neurips.cc/paper_files/paper/2010/file/d09bf41544a3365a46c9077ebb5e35c3-Paper.pdf}
}

@article{Pruessmann_Magn.Reson.Med._1999,
  title = {{{SENSE}}: Sensitivity Encoding for Fast {{MRI}}},
  shorttitle = {{{SENSE}}},
  author = {Pruessmann, K. P. and Weiger, M. and Scheidegger, M. B. and Boesiger, P.},
  date = {1999-11},
  journaltitle = {Magn. Reson. Med.},
  volume = {42},
  number = {5},
  eprint = {10542355},
  eprinttype = {pubmed},
  pages = {952--962},
  issn = {0740-3194},
  langid = {english}
}

@article{Robbins_Proc.ThirdBerkeleySymp.Math.Stat.Probab._1954,
  title = {An {{Empirical Bayes Approach}} to {{Statistics}}},
  author = {Robbins, Herbert},
  date = {1954},
  journaltitle = {Proceedings of the Third Berkeley Symposium on Mathematical Statistics and Probability},
  pages = {157--163},
  location = {1954},
  url = {http://digicoll.lib.berkeley.edu/record/112828}
}

@article{Roberts_MethodologyandComputinginAppliedProbability_2002,
  title = {Langevin {{Diffusions}} and {{Metropolis-Hastings Algorithms}}},
  author = {Roberts, G. O. and Stramer, O.},
  date = {2002-12},
  journaltitle = {Methodology and Computing in Applied Probability},
  volume = {4},
  number = {4},
  pages = {337--357},
  issn = {1387-5841, 1573-7713},
  doi = {10.1023/A:1023562417138},
  url = {https://link.springer.com/10.1023/A:1023562417138},
  urldate = {2025-11-23},
  langid = {english}
}

@article{Rosenzweig_Magn.Reson.Med._2019,
  title = {Simple Auto-Calibrated Gradient Delay Estimation from Few Spokes Using {{Radial Intersections}} ({{RING}})},
  author = {Rosenzweig, Sebastian and Holme, H. Christian M. and Uecker, Martin},
  date = {2019},
  journaltitle = {Magn. Reson. Med.},
  volume = {81},
  number = {3},
  pages = {1898--1906},
  issn = {1522-2594},
  doi = {10.1002/mrm.27506},
  url = {https://onlinelibrary.wiley.com/doi/abs/10.1002/mrm.27506},
  urldate = {2025-11-21},
  langid = {english}
}

@article{Scholand_Magn.Reson.Med._2025,
  title = {Rational Approximation of Golden Angles: {{Accelerated}} Reconstructions for Radial {{MRI}}},
  shorttitle = {Rational Approximation of Golden Angles},
  author = {Scholand, Nick and Schaten, Philip and Graf, Christina and Mackner, Daniel and Holme, H. Christian M. and Blumenthal, Moritz and Mao, Andrew and Assl\"ander, Jakob and Uecker, Martin},
  date = {2025},
  journaltitle = {Magn. Reson. Med.},
  volume = {93},
  number = {1},
  pages = {51--66},
  issn = {1522-2594},
  doi = {10.1002/mrm.30247},
  url = {https://onlinelibrary.wiley.com/doi/abs/10.1002/mrm.30247},
  urldate = {2025-11-21},
  langid = {english}
}

@inproceedings{Sohl-Dickstein_Proc.32ndInt.Conf.Mach.Learn._2015,
  title = {Deep {{Unsupervised Learning}} Using {{Nonequilibrium Thermodynamics}}},
  booktitle = {Proceedings of the 32nd {{International Conference}} on {{Machine Learning}}},
  author = {Sohl-Dickstein, Jascha and Weiss, Eric A and Maheswaranathan, Niru and Ganguli, Surya},
  date = {2015},
  series = {Proceedings of {{Machine Learning Research}}},
  volume = {37},
  pages = {2256--2265},
  publisher = {PMLR}
}

@inproceedings{Song_Adv.NeuralInf.Process.Syst._2019,
  title = {Generative {{Modeling}} by {{Estimating Gradients}} of the {{Data Distribution}}},
  booktitle = {Advances in {{Neural Information Processing Systems}}},
  author = {Song, Yang and Ermon, Stefano},
  date = {2019},
  volume = {32},
  publisher = {Curran Associates, Inc.},
  url = {https://proceedings.neurips.cc/paper_files/paper/2019/hash/3001ef257407d5a371a96dcd947c7d93-Abstract.html},
  urldate = {2024-01-16}
}

@inproceedings{Song_ICLR_2021,
  title = {Score-{{Based Generative Modeling}} through {{Stochastic Differential Equations}}},
  booktitle = {{{ICLR}} 2021: {{The Ninth International Conference}} on {{Learning Representations}}},
  author = {Song, Yang and Sohl-Dickstein, Jascha and Kingma, Diederik P. and Kumar, Abhishek and Ermon, Stefano and Poole, Ben},
  date = {2021},
  volume = {9},
  eprint = {2011.13456},
  eprinttype = {arXiv},
  eprintclass = {cs},
  doi = {10.48550/arXiv.2011.13456},
  url = {http://arxiv.org/abs/2011.13456},
  urldate = {2025-07-30},
  eventtitle = {International {{Conference}} on {{Learning Representations}}}
}

@article{Uecker_Magn.Reson.Med._2008,
  title = {Image Reconstruction by Regularized Nonlinear Inversion-Joint Estimation of Coil Sensitivities and Image Content},
  author = {Uecker, M. and Hohage, T. and Block, K. T. and Frahm, J.},
  date = {2008},
  journaltitle = {Magn. Reson. Med.},
  volume = {60},
  number = {3},
  pages = {674--682},
  publisher = {Wiley Online Library}
}

@article{Uecker_Magn.Reson.Med._2014,
  title = {{{ESPIRiT}}---an Eigenvalue Approach to Autocalibrating Parallel {{MRI}}: {{Where SENSE}} Meets {{GRAPPA}}},
  author = {Uecker, Martin and Lai, Peng and Murphy, Mark J. and Virtue, Patrick and Elad, Michael and Pauly, John M. and Vasanawala, Shreyas S. and Lustig, Michael},
  date = {2014},
  journaltitle = {Magn. Reson. Med.},
  volume = {71},
  number = {3},
  pages = {990--1001},
  publisher = {Wiley Online Library},
  doi = {10.1002/mrm.24751},
  langid = {english}
}

\clearpage

\appendix

\section*{Supporting Information}

\begin{figure}[htbp]
	\centering
	\includegraphics[width=\linewidth]{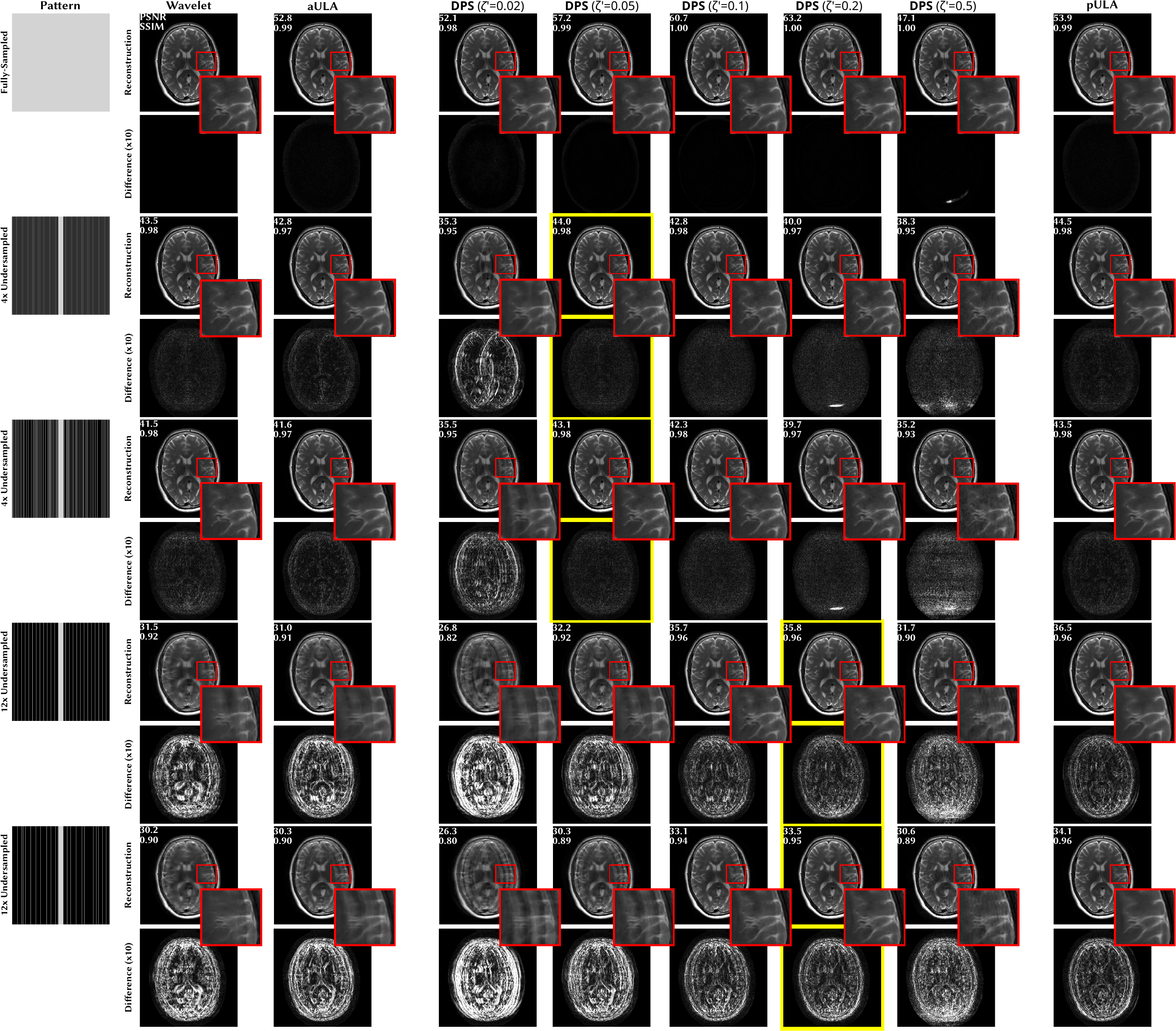}
	\caption{
		Comparison of different DPS likelihood weightings $\zeta'$ for the DPS method in Figure \ref{fig:comp}.
		The optimal weighting in terms of PSNR is highlighted in yellow. It depends on the acceleration factor.}
\end{figure}

\begin{figure}[htbp]
	\centering
	\includegraphics[width=0.8\linewidth]{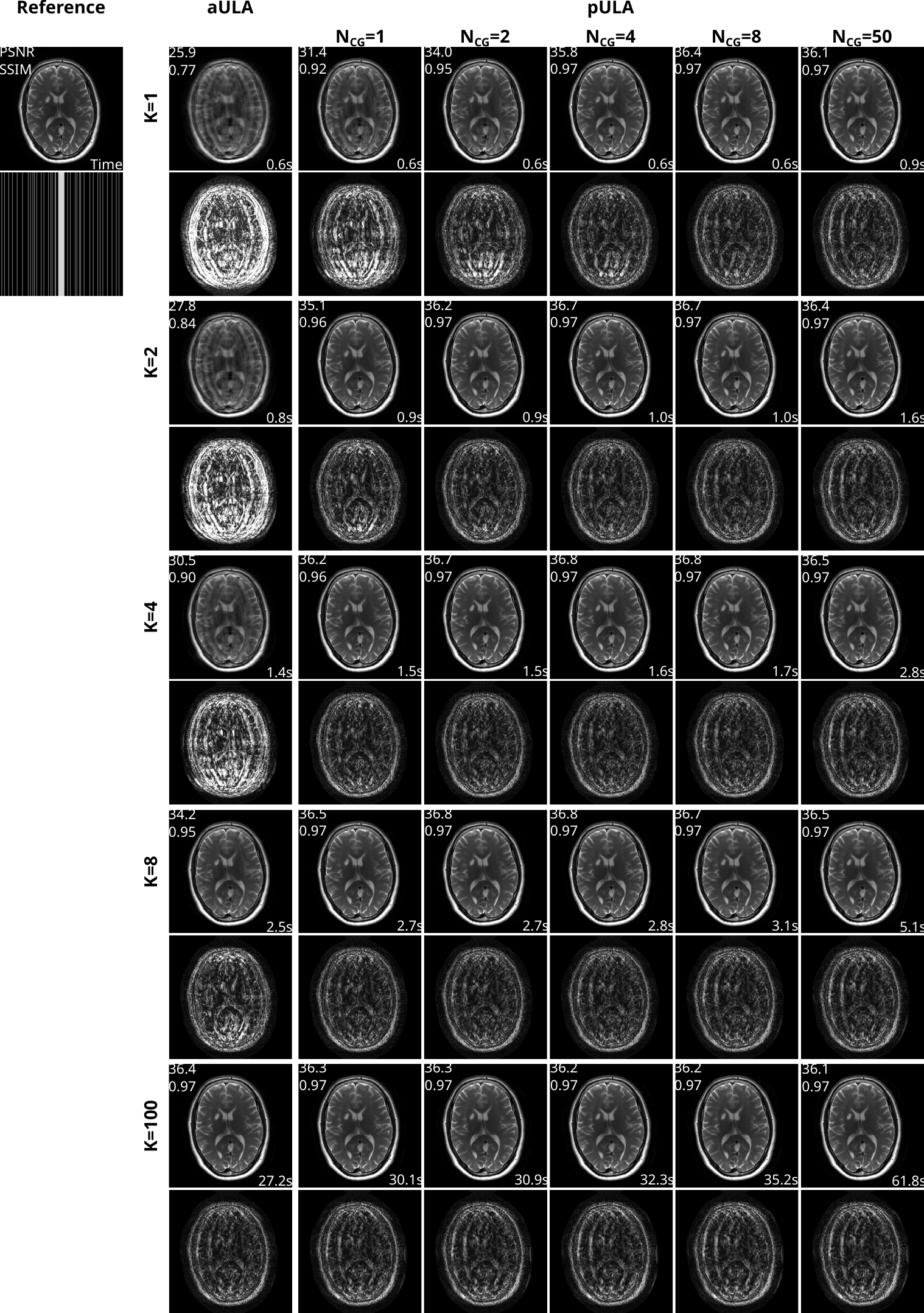}
	\caption{
		Comparison of annealed likelihood and exact likelihood with pULA for different numbers of Langevin iterations $K$ per noise level, and using different numbers of CG iterations $N_{CG}$.
		Reconstruction times are per sample.
		The quantitative metrics and intensity of the difference maps suggest that pULA with $K=1$ and $N_{CG}=8$ reaches similar reconstruction quality as the annealed likelihood with $K=100$
		and outperforms it for $K=8$, whereas the reconstruction time is significantly reduced by pULA.
		Further, increasing the number of CG iterations to $N_{CG}=50$ does not lead to visible changes in reconstruction quality, but slightly decreases the PSNR, potentially due to numerical inaccuracies.
	}
\end{figure}

\begin{figure}[htbp]
	\centering
	\includegraphics[width=\linewidth]{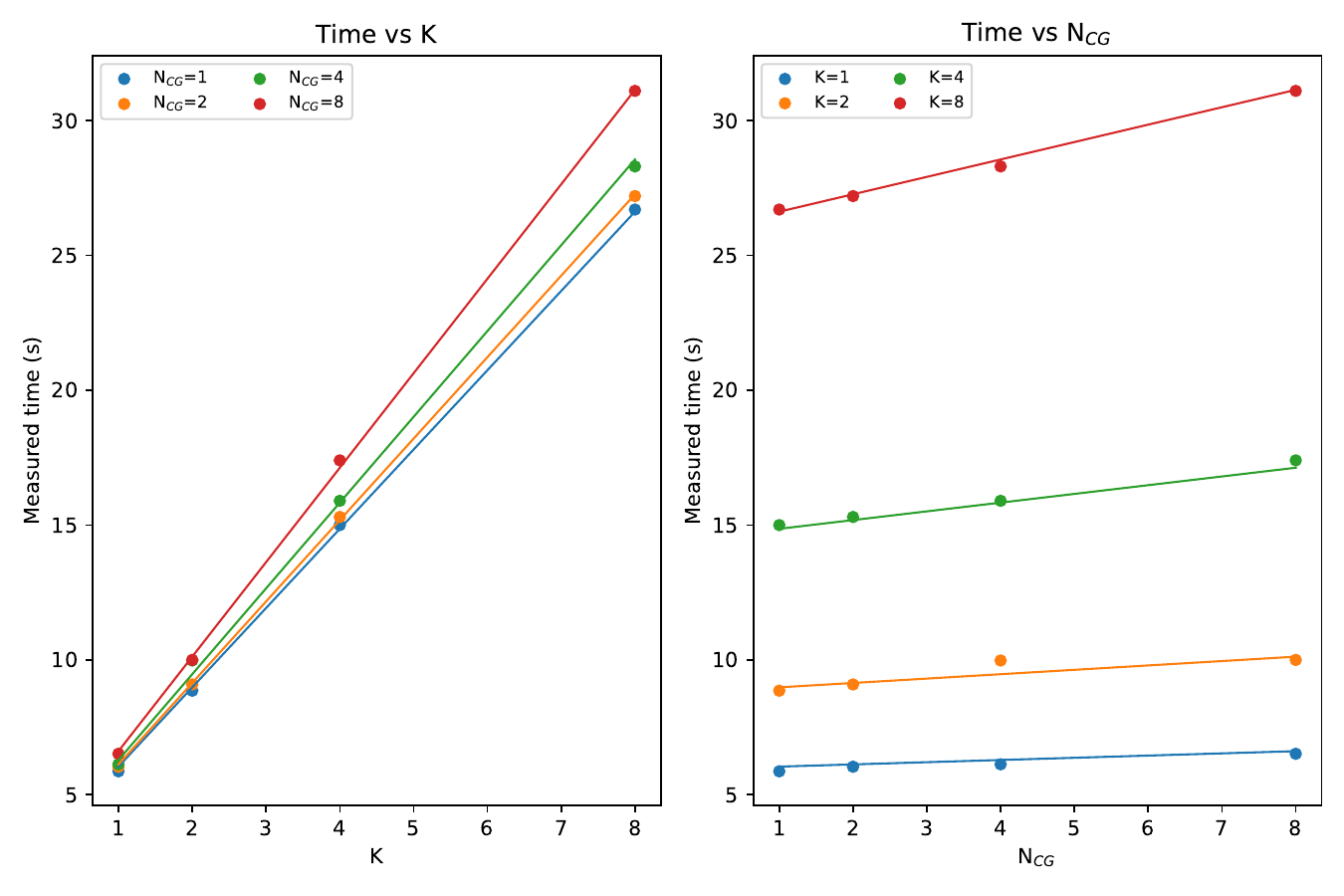}
	\caption{
		Measured reconstruction time of pULA depending on the number of CG and Langevin iterations per noise level.
		The measured times are presented as dots, whereas the solid lines represent linear fits using the model in \eqref{eq:time_model}.
		The fitted parameters are $t_{\mathrm{Network}}\approx\SI{4.5}{ms}$, $t_{A^HA}\approx\SI{0.13}{ms}$ and $t_{ini}\approx\SI{3.1}{s}$.
		The large $t_{\mathrm{ini}}$, is mostly due to initialization of the GPU and the network, but can be amortized when many reconstructions are performed.
		The evaluation of the neural network is the dominant cost, whereas the cost for applying $A^HA$ is about 34 times lower.
	}
\end{figure}

\begin{figure}[htbp]
	\centering
	\includegraphics[width=\linewidth]{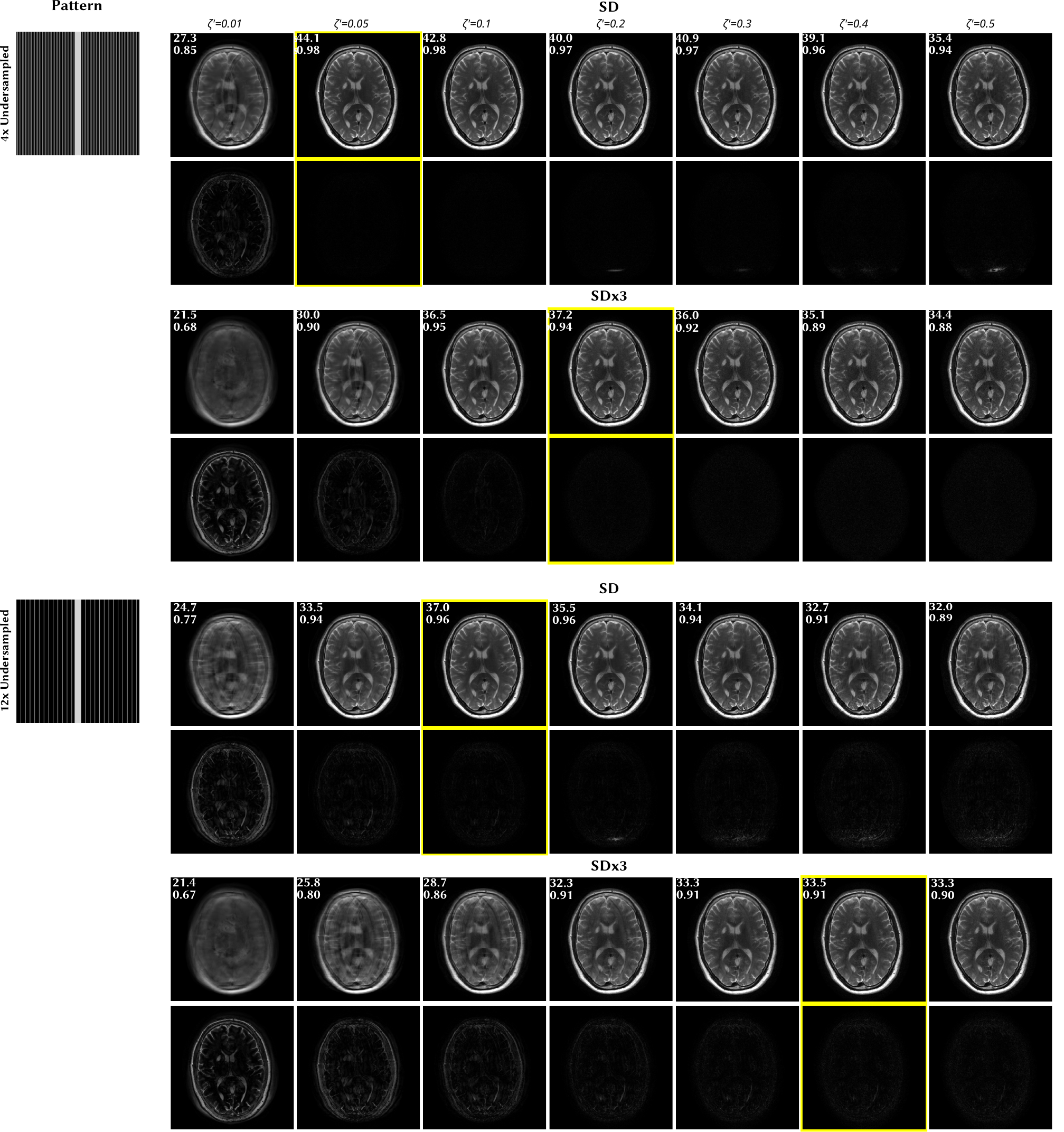}
	\caption{
		Comparison of different DPS likelihood weightings $\zeta'$ for the DPS method in Figure \ref{fig:noise_comp}.
		Highlighting that the optimal step size $\xi^\prime$ for different undersampling pattern and noise corruption.
	}
\end{figure}

\begin{figure}[htbp]
	\centering
	\includegraphics[width=\linewidth]{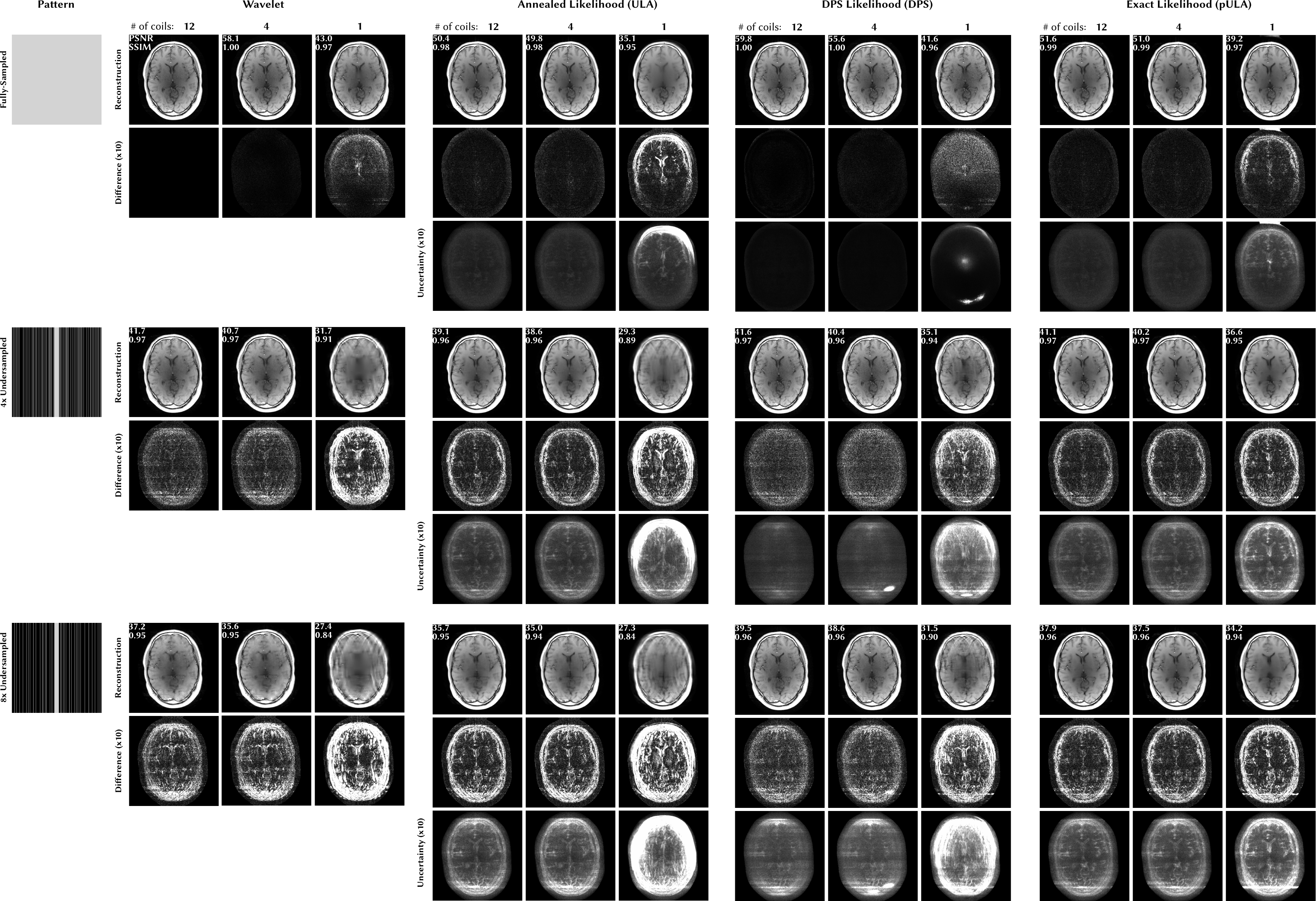}
	\caption{Reconstructions of a T1-weighted brain image for selected undersampling patterns and different numbers of virtual coils after coil compression. Uncertainty maps show the standard deviation over drawn samples.}
\end{figure}

\begin{figure}[htbp]
	\centering
	\includegraphics[width=\linewidth]{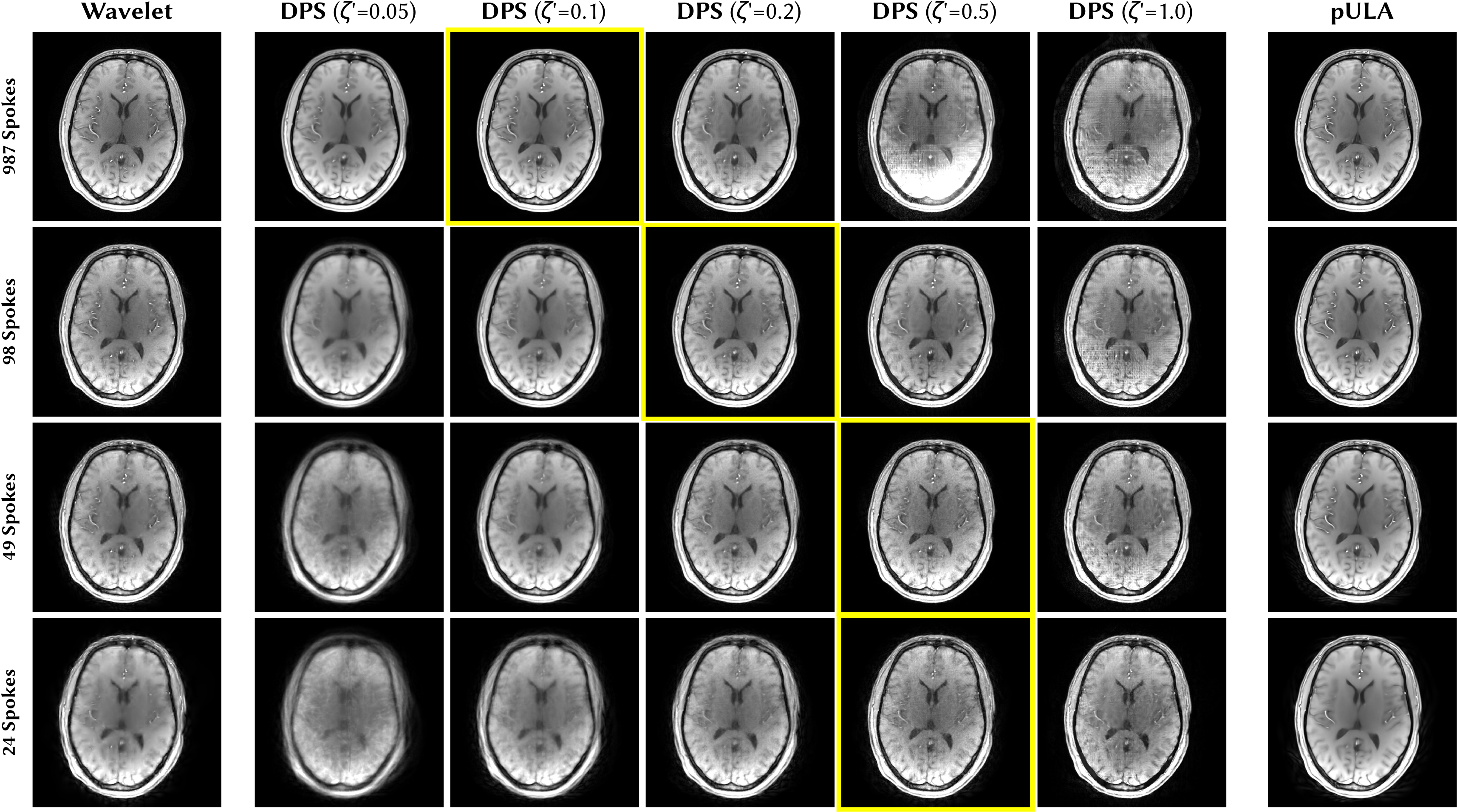}
	\caption{
		Comparison of different DPS likelihood weightings $\zeta'$ for the DPS method in Figure \ref{fig:noise_comp}.
		Highlighting that the optimal step size $\xi^\prime$ depends on the number of spokes.
		}
\end{figure}

\subsection*{S1 Exact Likelihood and Preconditioning Matrix in Variance Preserving Formulation}
In the variance exploding (VE) formulation of diffusion models, the smoothed prior distributions are defined by convolution with a Gaussian smoothing kernel, i.e.
\begin{align}
	p_t(x_t|x_0) = \mathcal{CN}(x_t; x_0, \sigma_t^2 I)\,.
\end{align}
In contrast, in the variance preserving (VP) formulation, the perturbation kernel reads
\begin{align}
	\tilde{p}_t(\tilde{x}_t|x_0) = \mathcal{CN}\left(\tilde{x}_t; \alpha_t x_0, (1-\alpha_t^2) I\right)\,,
\end{align}
where we us $\tilde{\cdot}$ to distinguish variables in the VP formulation from those in the VE formulation. Both formulations are equivalent by identifying
\begin{align}
	    x_t = \frac{\tilde{x}_t}{\alpha_t}&&\text{and}&& \sigma_t^2 = \frac{1-\alpha_t^2}{\alpha_t^2}\,.
\end{align}
%Using the chain rule, the reparameterized prior score function in the VP formulation reads:
%\begin{align}
%    \nabla_{x_t} \log p_t(x_t) = \alpha_t \nabla_{\tilde{x}_t} \log \tilde{p}_t(\tilde{x}_t) \,.
%\end{align}
%The ULA update step in the VP formulation for posterior sampling is then given by
%\begin{align}
%    \mathbf{\tilde{x}}_t^{k+1}=\mathbf{\tilde{x}}_t^k+\gamma
%    \nabla_{\bar{\mathbf{\tilde{x}}}}\left[\log \tilde{p}(\mathbf{y}|\mathbf{\tilde{x}}_t^k) + \log \tilde{p}_t(\mathbf{\tilde{x}}_t^k)\right]+\sqrt{2\gamma}\mathbf{z}^k&&\mathbf{z}^k\sim\mathbb{C}\mathcal{N}(\mathbf{0},\mathbb{I})\,,
%\end{align}
The \textit{Exact Likelihood} term from \eqref{eq:diff_lik_annelead} reads in the VP formulation
\begin{align}
    \tilde{p}^{\mathrm{exact}}(\mathbf y | \mathbf{\tilde{x}}_t) \propto \exp\left(-\lVert\mathbf{y}-\alpha_t^{-1}A\mathbf{\tilde{x}}_t\rVert_2^2\right),
    \label{eq:diff_lik_annelead_VP}
\end{align}
yielding a ULA update step of
\begin{align}
    \mathbf{\tilde{x}}_t^{k+1}=\mathbf{\tilde{x}}_t^k+\gamma
    \left[\alpha_t^{-1}A^H(\mathbf{y}-\alpha_t^{-1}A\mathbf{\tilde{x}}_t^k) + \nabla_{\bar{\tilde{\mathbf{x}}}}\log \tilde{p}_t(\mathbf{\tilde{x}}_t^k)\right]+\sqrt{2\gamma}\mathbf{z}^k&&\mathbf{z}^k\sim\mathbb{C}\mathcal{N}(\mathbf{0},\mathbb{I})\,.
\end{align}
By reparametrization the Hessian of the log-prior transforms with $\alpha_t^{2}$ and, hence, the inverse preconditioning matrix in the VP formulation is given by
\begin{align}
    \tilde{M}_t^{-1} = \alpha_t^{-2}A^HA + \alpha_t^{-2}\sigma_t^{-2} \mathbb{I} = \alpha_t^{-2}A^HA + \left(1 - \alpha_t^2\right)^{-1} \mathbb{I}\,.
\end{align}

\subsection*{S2 Warmstart of Conjugate Gradient Algorithm}

To reduce the number of CG iterations when applying the preconditioning matrix, we warmstart the CG algorithm. For this we first note that the pULA update as expressed in the main manuscript can be rewritten as
\begin{align}
\bvec{x}_t^{k+1} &= \bvec{x}_t^k + \gamma {M}_{t}
\left[
	A^H\left(\bvec{y} + \sqrt{\frac{2}{\gamma}}\bvec{n}^k_1 - A\bvec{x}_t^k\right) + \nabla_{\bar{\mathbf{x}}} \log p(\bvec{x}_t^k) + \sqrt{\frac{2}{\gamma\sigma^2_t}}\bvec{n}^k_2
\right]\notag\\
&= \bvec{x}_t^k + {M}_{t}
\left[
	A^H\left(\gamma\bvec{y} + \sqrt{2\gamma}\bvec{n}^k_1 - A\gamma\bvec{x}_t^k\right) + \gamma\nabla_{\bar{\mathbf{x}}} \log p(\bvec{x}_t^k) + \sqrt{\frac{2\gamma}{\sigma^2_t}}\bvec{n}^k_2
\right]\notag\\
&= {M}_{t}
\left[
	A^H\left(\gamma\bvec{y} + \sqrt{2\gamma}\bvec{n}^k_1 - A\gamma\bvec{x}_t^k\right) + \gamma\nabla_{\bar{\mathbf{x}}} \log p(\bvec{x}_t^k) + \sqrt{\frac{2\gamma}{\sigma^2_t}}\bvec{n}^k_2
	+ A^HA\bvec{x}_t^k + \sigma_t^{-2}\bvec{x}_t^k
\right]\,.
\end{align}
In this form, the pULA update is directly given by the solution of the CG algorithm.
We initialize this conjugate gradient algorithm with
\begin{align}
	\bvec{x}_{t,\mathrm{init}}^{k+1} = \bvec{x}_t^k + \gamma \sigma_t^2 \nabla_{\bar{\mathbf{x}}} \log p(\bvec{x}_t^k) + \sqrt{2\gamma\sigma_t^2}\bvec{n}^k_2\,,
\end{align}
corresponding to an update step only using the prior score. Supporting Figure S2 shows, that using this warmstart strategy, already one CG iteration leads to improved reconstructions compared to the annealed likelihood method.

\subsection*{S3 Implementation of DPS in the Variance Exploding Formulation}
The DPS \cite{Chung_Elev.Int.Conf.Learn.Represent._2023} method approximates the diffused likelihood by
\begin{align}
    p^{\mathrm{DPS}}(\mathbf y | \mathbf{ x}_t) \propto \exp\left(-\lVert\mathbf{y}-A\mathbb E[\mathbf{ x}_0|\mathbf{ x}_t]\rVert_2^2\right).
    \label{eq:diff_lik_dps_sup}
\end{align}
We have implemented DPS in the variance exploding formulation by integrating the corresponding reverse stochastic differential equation using the predictor approach described in Algorithm 2 of \cite{Song_ICLR_2021}. Using $N$ steps to discretize the time $t$ and denoting $\sigma_i = \sigma(i/N)$ the respective noise levels, DPS likelihood from \eqref{eq:diff_lik_dps_sup} into the predictor update step yields
\begin{align}
    \mathbf{x}_{i} = \mathbf{x}_{i + 1} &+ \left(\sigma_{i+1}^2-\sigma_i^2\right) \left[\nabla_{\bar{\mathbf{x}}} \log p(\mathbf{x}_{i+1}) + \nabla_{\bar{\mathbf{x}}} \log p^{\mathrm{DPS}}(\mathbf{y}|\mathbf{x}_{i+1})\right] \notag\\
     & + \sqrt{\sigma_{i+1}^2-\sigma_i^2}\mathbf{z}_i&&\mathbf{z}_i\sim\mathbb{C}\mathcal{N}(\mathbf{0},\mathbb{I})\,.
\end{align}
It was observed by \citeauthor{Chung_Elev.Int.Conf.Learn.Represent._2023} \cite{Chung_Elev.Int.Conf.Learn.Represent._2023} that using equal weighting of the likelihood and the prior score leads to suboptimal results, such that they introduced a weighting parameter $\zeta$ to balance the two terms. Heuristically, the weighting was chosen based on the current residual. We set it to
\begin{align}
	\zeta_{i+1} =
	\frac{\zeta'}{\left(\sigma_{i+1}^2-\sigma_i^2\right)\lVert\mathbf{y}-A\mathbb E[\mathbf{ x}_0|\mathbf{ x}_{i+1}]\rVert }\,,
\end{align}
where $\zeta'$ is a hyperparameter that needs to be tuned. The implemented update step for our DPS implementation reads
\begin{align}
    \mathbf{x}_{i} = \mathbf{x}_{i + 1} &+ \left(\sigma_{i+1}^2-\sigma_i^2\right) \left[\nabla_{\bar{\mathbf{x}}} \log p(\mathbf{x}_{i+1}) + \zeta_{i+1} \nabla_{\bar{\mathbf{x}}} \log p^{\mathrm{DPS}}(\mathbf{y}|\mathbf{x}_{i+1})\right] \notag\\
     & + \sqrt{\sigma_{i+1}^2-\sigma_i^2}\mathbf{z}_i&&\mathbf{z}_i\sim\mathbb{C}\mathcal{N}(\mathbf{0},\mathbb{I})\,.
\end{align}

\subsection*{S4 Runtime Analysis of aULA and pULA}
\subsubsection*{Methods}

The computational complexity of pULA was investigated by varying the number of Langevin iterations $K=1,2,4,8$ and CG iterations $N_{CG}=1,2,4,8$.
The expected dominant computational cost of aULA are the score network evaluation ($t_{\mathrm{Network}}$) and the normal operator ($t_{A^HA}$), once in each Langevin iteration.
For pULA, $N_{CG} + 1$ additional applications of $A^HA$ are required in each Langevin iteration for the preconditioning and $N_{CG} + 1$ evaluations for the initialization.
Hence, the total reconstruction time $t_{\mathrm{total}}$ is modeled as
\begin{align}
	    t_{\mathrm{total}} = N_S N  K \cdot t_{\mathrm{Network}} + N_S N K (N_{CG} + 1) \cdot t_{A^HA} + (N_{CG} + 1) \cdot t_{A^HA} + t_{\mathrm{init}}\,,
	    \label{eq:time_model}
\end{align}
where $t_{\mathrm{init}}$ summarizes all additional one-time computations such as initialization of the GPU and the score network.
We fitted this model to the measured reconstruction times using least squares to estimate $t_{\mathrm{Network}}$, $t_{A^HA}$, and $t_{\mathrm{init}}$.

\subsubsection*{Results}

Reconstruction results for different numbers of Langevin and CG iterations are shown in Supporting Figure S2.
It shows that by using preconditioning, the number of expensive network applications can be
reduced without loss in reconstruction quality, whereas increasing the number of CG iterations
beyond 8 does not lead to further significant improvements.

The measured reconstruction times as functions of the number of Langevin and CG iterations are shown in Supporting Figure S3,
together with the fitted model from \eqref{eq:time_model}.
The fitted parameters are $t_{\mathrm{Network}}\approx\SI{4.5}{ms}$, $t_{A^HA}\approx\SI{0.13}{ms}$ and $t_{ini}\approx\SI{3.1}{s}$.
It confirms that in our setting the cost
for applying the network is the dominant cost
whereas applying $A^HA$ is about 34 times lower.

%TC:endignore

\end{document}